\documentclass{article}
\usepackage[final]{pdfpages} 
\usepackage{longtable}
\usepackage{multicol}

\usepackage{tabularx}
\usepackage{ gensymb }
\usepackage{ graphicx ,float}
\usepackage{ upgreek }
\usepackage { amssymb }
\usepackage { amsmath }
\usepackage { multirow }
\usepackage {placeins}
\usepackage{epstopdf}
\usepackage { amsthm }
\usepackage{geometry}
\usepackage[cyr]{aeguill}
\usepackage{placeins,hyperref}
\usepackage[utf8]{inputenc}

\newcommand{\vep}{\varepsilon}
\renewcommand{\geq}{\geqslant}
\usepackage{algorithm}
\usepackage{algorithmic}

\usepackage{color}
\usepackage{xpatch}

\newcommand{\pen}{\text{pen}}

\renewcommand{\leq}{\leqslant}
\renewcommand{\geq}{\geqslant}

\newcommand{\bs}{\boldsymbol}

\usepackage{authblk}
\title{AcSel: selecting variables with accuracy in correlated data sets}
\author[1,2]{Nicolas Jung \thanks{Corresponding author, please write at : {nicolas.julien.jung@gmail.com}}}
\author[1]{Frédéric Bertrand}
\author[1]{Myriam Maumy-Bertrand}
\affil[1]{Institut de Recherche en Mathématique Avancée, UMR 7501}
\affil[2]{Laboratoire d’Immunogénétique Moléculaire Humaine, Institut National de la Santé et de la Recherche Médicale, Unité Mixte de Recherche S1109}

\begin{document}

\maketitle

\begin{abstract}
With the emergence high-throughput technologies, it is possible to measure large amounts of data relatively low cost.  Such situations arise in many fields from sciences to humanities, and variable selection may be of great help to answer challenges that are specific to each of them. Variable selections may allow to know, among all measured variables, which are of interest and which are not. A lot of methods have been proposed to handle this issue, with the Lasso and other penalized regression as special cases. These methods fail in some cases and linear correlation between explanatory variables is the most common, especially in big datasets. In this article, we introduce AcSel, a wrapping algorithm to enhance the accuracy of any variable selection method.
\end{abstract}

\section{Introduction}

\sloppy

The problem of variable selection has received an increasing attention over the last years \cite{fan_statistical_2006} and is one of the most important challenges for the 21st century \cite{donoho2000high}. Indeed, technological innovations make it possible to measure large amounts of data  relatively low cost. As a consequence, problems in which the number $P$ of variables is greater that the number $N$ of observations have become common. As reviewed by Fan and Li \cite{fan_statistical_2006}, such situations arise in many fields from sciences to humanities, and variable selection may be of great help to answer challenges that are specific to each of them. For example, in biology, thousands of messenger RNA (mRNA) gene expressions \cite{lipshutz1999high}  may be potential predictors of some illness. Other examples are imagery (magnetic resonance image, nuclear magnetic resonance, satellite images...), financial engineering and risk management, or health studies \cite{fan_statistical_2006}. Moreover, in such studies, the correlation between variables is often very strong \cite{segal2003regression} and variable selection methods often fail to choose the informative variables among those which are not.\\

In this article, we assume that our data is generated by a multivariate linear model: 

\begin{equation}\label{model}
\bs{y} = \mu \bs{1}_N + \bs{X}\bs{\beta} + \bs{\vep},
\end{equation}

where $\bs{y} = (y_1,...,y_N)'$ is the response variable, $\mu$ is the mean variable response, $\bs{1}_N$ is a vector of length $N$ containing only ones, $\bs{X}=(\bs{x}_{1.},...,\bs{x}_{P.})$ is the design matrix of size $N \times P$, $N>2$,  with $\bs{x}_{p.}=(x_{p1},...,x_{pN})'$ which are the variables and $\bs{\vep} = (\vep_1,...,\vep_N)$ is a Gaussian noise vector which is the realization of some random law with a mean of 0 and an  unknown variance $\sigma^2$. Furthermore, we will assume that the   vector of parameters $\bs{\beta} = (\beta_1,...,\beta_P)'$ is sparse. In other words, we will assume that $\beta_i=0$ except for a quite small proportion of elements of the vector. We note $\mathcal{S}$ as the set of indexes for which $\beta_i \neq 0$ and $q<\infty$ is the cardinal of this set $\mathcal{S}$. Without any loss of generality, we will assume that $\beta_p \neq 0$ if and only if $p \leqslant q$. Moreover, we assume that the response and the  variables are centred  and that $\|\bs{x}_p.\|^2=1$ for $p=1,...,P$ where $\|\cdot\|$ stands for the usual euclidean norm; in this context, we have $\mu= 0$. \\

When dealing with a problem of variable selection, there are three main goals. We enumerate them in increasing level of difficulty:

\begin{enumerate}
\item The prediction goal, in which you want $\bs{\hat{y}}$ to be as close as possible to $\bs{y}$.
\item The estimation goal, in which you want $\bs{\hat{\beta}}$ to be as close as possible to $\bs{\beta}$.
\item The estimation of the support,  in which you want  $\mathbb{P}(\mathcal{S} = \mathcal{\hat{S}})$ to be close to one.
\end{enumerate}

Fan and Li \cite{fan2001variable} proposed another desirable property, the oracle property, which combines goals 2 and 3. Precisely, a method is said to have the oracle property if it discovers the correct support, and if the rate of convergence of $\bs{\hat{\beta}}$ toward $\bs{\beta}$ is optimal (\textit{i.e.} the same as in the case in which the correct support is known). Here, our interest is mainly in the third goal, \textit{i.e.} in identifying the correct support $\mathcal{S}$. This kind of issue arises in many fields, for example in biology, where it is of greatest interest to discover which specific molecules are involved in a disease \cite{fan_statistical_2006}.  \\

There is a vast literature dealing with the problem of variable selection in both statistical and machine learning areas (\cite{fan_statistical_2006,fan2010selective}).  The main variable selection methods can be gathered in the common framework of penalized likelihood. The estimate $\hat{\bs{\beta}}$ is then given by:

\begin{equation}\label{common}
\hat{\bs{\beta}} = \text{arg}\min_{\bs \beta \in \mathbb{R}^P} \left[- {\ell_N ({\bs{\beta}} ) }  +  \sum_{p=1}^P \pen_{\bs\lambda}({\beta_p})\right],
\end{equation}

where $\ell_N (.)$ is the log-likelihood function,  $ \pen_{\bs\lambda}(.)$ is a penalty function with $k$ parameters and $\bs{\lambda} = (\lambda,\lambda_2,\lambda_3,...,\lambda_k)'$. As the goal is to obtain a sparse estimation of the vector of parameters $\bs \beta$, a natural choice for the penalty function is to use the so-called $L_0$ norm ($\|.\|_0$) which corresponds to the number of non-vanishing elements of a vector:

\begin{equation}
\begin{array}{lcccc}  
 \pen_{\bs\lambda} &:&  \mathbb{R} & \to &\{0,\lambda\}\\
    &&x &\mapsto&\left\{
\begin{array}{c l}     
   \pen_{\bs\lambda} (x)= \lambda & \text{if } x\neq 0\\
   \pen_{\bs\lambda} (x)= 0 &\text{else}
\end{array}\right.
\end{array}  \Rightarrow \sum_{p=1}^P \pen_{\bs\lambda}({\beta_p}) = \lambda \| {\bs{\beta}}\|_0\cdot
\end{equation}

For example, when $\lambda = 1$, we get the Akaike Information Criterion (AIC) \cite{akaike1974new} and when  $\lambda = \frac{\log(N)}{2}$ we get the Bayesian Information Criterion (BIC) \cite{schwarz1978estimating}. Another slightly different formulation leads to Mallow's $C_p$ \cite{mallows1973some} or to the Risk Inflation Criterion \cite{foster1994risk}. In the context of Gaussian independent and identically distributed (i.i.d.) errors in the model described in equation (\ref{model}), the following holds \cite{burnham2002model}:

\begin{equation}
-\ell_N ({\bs{\beta}} ) = \frac{N}{2}\log \left( \frac{\| \bs{y}-\bs X \hat{\bs \beta} \|^2}{N} \right) + K_1,
\end{equation}

where $K_1$ is a constant. Up to an affine transformation of the log-likelihood \cite{fan2010selective}, we see that equation (\ref{common}) is equivalent to: 

\begin{equation}\label{other}
\hat{\bs{\beta}} = \text{arg}\min_{\bs \beta \in \mathbb{R}^P}\left[{\| \bs{y}-\bs X {\bs \beta} \|^2} +\sum_{p=1}^P \pen_{\bs\lambda}({\beta_p}) \right].
\end{equation}

A lot of different penalties can be found in the literature. Solving this problem with  $\|.\|_0$ as part of the penalty is an NP-hard problem \cite{natarajan1995sparse,fan2010selective}. It cannot be used in practice when $P$ becomes large, even when it is employed with some search strategy like forward regression, stepwise regression\cite{hocking1976biometrics}, genetic algorithms \cite{koza1999genetic}... Donoho and Elad \cite{donoho2003optimally} show that relaxing  $\|.\|_0$  to norm $\|.\|_1$ ends, under some assumptions, to the same estimation. This result encourages the use of a wide range of penalty based on different norms. For example, the case where $\pen_{\bs\lambda}({\beta_p}) = \lambda |\beta_p| $ is the Lasso estimator \cite{tibshirani1996regression} (or equivalently Basis Pursuit Denoising \cite{chen2001atomic})  whereas $\pen_{\bs\lambda}({\beta_p}) = \lambda \beta_p^2$ leads to the Ridge estimator \cite{hoerl1970ridge}. These two last cases can be seen as a special case of Bridge regression \cite{frank1993statistical} in which $\pen_{\bs\lambda}({\beta_p})  = \lambda |\beta_p|^b$ with $0<b\leq 2$. Nevertheless, the penalty term induces variable selection only if:

$$
\min_{x \geqslant 0}\left( \frac{\text{d}\pen_{\bs\lambda}(x)}{\text{d}x}+x \right) >0.
$$

 This explains why the Lasso regression allows variable selection while the Ridge regression does not. As it is well known \cite{zou2006adaptive}, the Lasso leads to a biased estimate. The SCAD (smoothly clipped absolute deviation) \cite{fan1997comments}, MCP (minimax concave penalty) \cite{zhang2010nearly} or adaptative Lasso \cite{zou2006adaptive} penalties all address this problem. The popularity of such variable selection methods is linked to  fast algorithms including LARS (least-angle regression) \cite{efron2004least}, coordinate descent \cite{wu2008coordinate} or PLUS \cite{zhang2010nearly}. \\%However, it is well known that the Lasso estimator allows variable selection whereas the Ridge estimator does not. According to Fan and Li \cite{fan2001variable}, the penalty function should have the three following properties:
%
%\begin{enumerate}
%\item Sparsity: the estimator should set to $0$ all ``small'' coefficients, resulting in a dimension reduction. This holds when: $\min_{x \geqslant 0}\left( \frac{\text{d}\pen_{\bs\theta}(x)}{\text{d}x}+x \right) >0.$
%\item Unbiasedness: the estimator should be nearly unbiased. In particular, large coefficients should not be affected by the penalty in an important way. This holds when: $\frac{\text{d}\pen_{\bs\theta}(x)}{\text{d}x} =0$ when $|x|$ becomes large.
%\item Continuity: the estimator should be continuous in order to reduce instability in model selection.  This holds when: $
%\text{arg}\min_{x \geqslant 0}\left( \frac{\text{d}\pen_{\bs\theta}(x)}{\text{d}x}+x \right) =0.$
%\end{enumerate}
%
%Although the Lasso estimator satisfies the first property, this estimator results in a biased estimate. For example, if X has an orthogonal design, the Lasso (equation (\ref{lasso})) reduces to: $\hat{\bs \beta}^{Lasso} = \text{sgn}\left(\hat{\bs \beta}^{OLS}\right) \left[\left|\hat{\bs \beta}^{OLS}\right|- \lambda \right]_+$, where $\hat{\bs \beta}^{OLS}$ is the ordinary least square estimation, $  \text{sgn}(.)$ is the function that maps all positive numbers to 1, all negative numbers to -1 and 0 to 0 while $[]_+$ is the notation for the positive truncation.

Nevertheless, the goal of identifying the correct support of the regression is complicated and the reason why variable selection methods fail to select the  set of non-zero variables $\mathcal{S}$ can be summed up in one word: linear correlation. Choosing the Lasso regression as a special case, Zhao and Yu \cite{zhao2006model} (and simultaneously Zou \cite{zou_adaptive_2006}) found an almost necessary and sufficient condition for Lasso sign consistency (\textit{i.e.} selecting the  non-zero variables with the correct sign). This condition is known as ``irrepresentable condition'': 

\begin{equation}
\left| \bs{X}'_{\setminus\mathcal{S}}\bs{X}_{\mathcal{S}} \left(  \bs{X}'_{\mathcal{S}}\bs{X}_{\mathcal{S}}\right)^{-1}   \text{sgn}(\bs{\beta}_\mathcal{S})\right|< \bs{1},
\end{equation}

where $\bs{X}_\mathcal{S} = (x_{ij})_{i,j \in \mathcal{S}^2}$, $\bs{\beta}_\mathcal{S} = ({\beta_p})_{p \in \mathcal{S}}$. In other words, when $ \text{sgn}(\bs{\beta}_\mathcal{S})=\bs{1}_q$, this can be seen as the regression of each variable which is not in $\mathcal{S}$ over the variables which are in $\mathcal{S}$. As all variables in the matrix $\bs X$ are centred, the absolute sum of the regression parameters should be smaller than 1 to satisfy this ``irrepresentable condition''. \\

%Although $ \text{sgn}(\bs{\beta}_\mathcal{S})$ is unknown in practice, Zhao and Yu \cite{zhao2006model}  give five corollaries in which the ``irrepresentable condition'' holds in practice. For example, if $\bs{\beta}$ has $q$ non-zero entries (\textit{i.e.} $q$ is the cardinal of $\mathcal{S}$) and the correlation matrix $C$ has the following form:
%
%$$
%C=\begin{pmatrix}
%   1 &r&\cdots&\cdots & r \\
%    r &1&\ddots& & \vdots \\
%\vdots&\ddots&\ddots&\ddots & \vdots\\
%  \vdots &&\ddots&1 & r\\
%   r &\cdots&\cdots&r & 1
%\end{pmatrix},
%$$
%
%with $0<r<\frac{1}{2q-1}$, then the ``irrepresentable condition'' holds. As shown in Figure 1, this condition becomes  far too restrictive when $q>2$, in particular in $P>>N$ designs. \\
%
%\begin{figure}[h]
%\center
%\includegraphics[width=5cm]{exemple1.pdf}
%\caption{Evolution of the bound of correlation in function of $q$.}
%\end{figure}

Facing this issue, existing variable selection methods can be split into two categories:

\begin{itemize}
\item Those which are ``regularized'' and try to give a similar coefficients to variables which are correlated (\textit{e.g.}: elastic net \cite{zou2005regularization}), 
\item Those which are not ``regularized'' and pic up one variable among a set of correlated variables (\textit{e.g.}: the Lasso  \cite{tibshirani1996regression}).
\end{itemize}

 The former group can then be split into methods in which groups of correlation are known,  such as the group Lasso \cite{yuan2006model,friedman2010note} and those in which groups are not known  as in  the elastic net \cite{zou2005regularization}. The latter  combines the $\mathcal{L}_1$ and the $\mathcal{L}_2$ norm and takes advantage of both. Broadly speaking, non-regularized methods will select some co-variables among a group of correlated variables while regularized methods will select all variables in the same group with similar coefficients (see example in Figure 2). \\

\begin{figure}[h]
\center
\includegraphics[width=\textwidth]{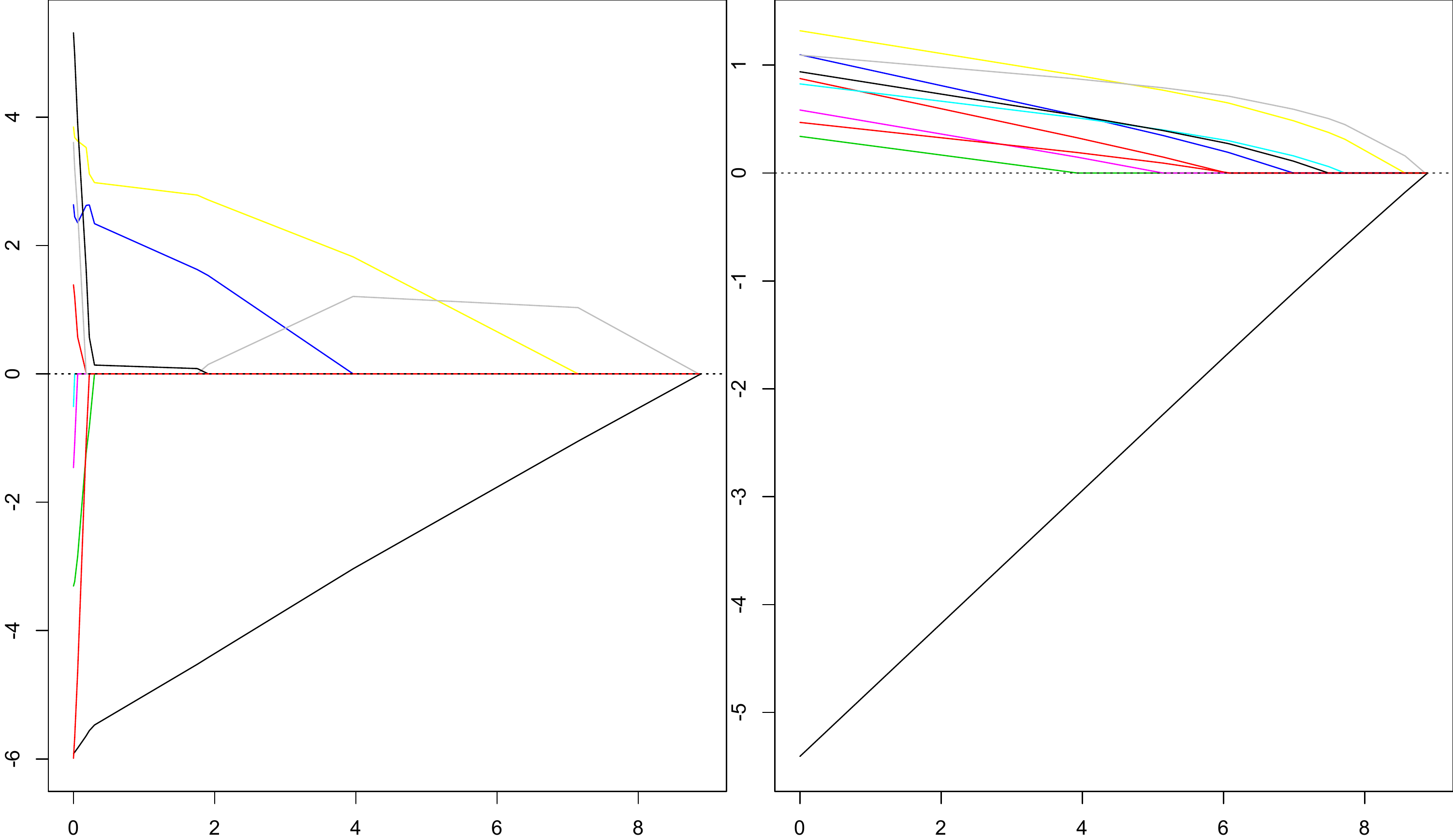}
\caption{In this example $N=20, P = 10$, $\bs\beta = (1,1,0,...,0)'$. The mean correlation between $\bs{x}_{1.}$ and the other variables is $0.20$ while the mean correlation between all the other variables is $0.95$. The x-axis corresponds to the value of the penalty parameter $\lambda$ ; the greater the parameter, the stronger the constraint. Right: with the lasso regression, no regularization is made. Left: with the elastic net regression, the coefficients of correlated variable are similar.}
\end{figure}

However, none of these selection methods distinguishes between variables that were selected for inclusion in the model with confidence and those that were not. In this article, we propose the AcSel algorithm  that can provide a confidence factor for selected variables. Our new algorithm will be useful in different contexts, including biology where it will allow high precision selection of relevant therapeutic targets.\\

The rest of this article is organized as follows. In section 2 we present our new algorithm, in section 3 we drive some simulation studies. A real dataset will be analysed in section 4, while section 5 will end with some remarks and conclusion notes. 

\section{Methods}

The AcSel algorithm has been designed in a general framework whose goal is to enhance the abilities of any variable selection method, especially those which are not regularized\footnote{Regularized should be understand as it is defined in this article, namely, a regularized method should give similar estimations for coefficients if the corresponding variables are strongly correlated.}. The main goal of this algorithm is to improve the precision, \textit{i.e.} the proportion of selected variables which really are in $\mathcal{S}$.

%Additionally we want to improve the recall, \textit{i.e.} the proportion of variables in $\mathcal{S}$ which have been selected. These two objectives are obviously antagonistic, and it has been pointed out \cite{wu2007controlling} that  over-fitting is linked with the selection of uninformative variables and that under-fitting (variables which are not in $\mathcal{S}$) is linked exclusion of the informative ones (variables which are  in $\mathcal{S}$). This is why we want our algorithm to return two distinct lists: the list of variables selected with high confidence and the list of all variables that are possibly informative. \\

\subsection{Introduction}

The main idea of our algorithm is to consider that groups of variables of  the matrix $\bs{X}$ which are linearly correlated are independent realizations of the same random function. According to this random function, correlated variables are then perturbed.  Strictly speaking,  the use of  noise to make the difference between the informative and the uninformative variables  is not a new idea. For example, it has been shown that adding random pseudo-variables decreases over-fitting \cite{wu2007controlling}. In  the case where $P>N$ the pseudo-variables are generated either with independent Gaussian laws $\mathcal{N}(0,1)$ or by using permutations on the matrix $\bs X$ \cite{wu2007controlling}. Another approach consists in adding noise to the response variable and leads to similar results \cite{luo2006tuning}. The rational of this last method is based on the work of Cook and Stefanski \cite{cook1994simulation} which introduces the  simulation-based algorithm SIMEX \cite{cook1994simulation}. Adding noise to the matrix $\bs{X}$ has already been used in the context of microarrays \cite{chen2007noise}. Simsel \cite{eklund2012simsel} is an algorithm that both adds noise to variables and uses random pseudo-variables.  One new and interesting approach is stability selection \cite{meinshausen2010stability}  in which the variable selection method is applied on sub-samples, and informative variables are defined as variables which have a high probability of being selected. Bootstraping has been applied to the Lasso on both response variable and the  matrix $\bs{X}$ with better results in the former case \cite{bach2008bolasso}.  The random Lasso, in which variables are weighted with random weights, has also been introduced \cite{wang2011random}.\\

In this article,  following the idea of using simulation to enhance the variable selection methods, we propose the AcSel algorithm. Unlike other algorithms reviewed above, our algorithm takes care of the correlation structure of the data. Furthermore, our algorithm is motivated by the fact that in the case of non-regularized variable selection methods, if a group contains variables that are highly correlated together, one of them will be chosen ``at random'' \cite{zou2005regularization}. \\

As we assume that  the variables are centred  and that $\|\bs{x}_{p.}\|^2=1$ for $p=1,...,P$, we know that $\bs{x}_{p.} \in \mathcal{S}^{N-2}$. Indeed, the normalization puts the variables on the unit sphere $\mathcal{S}^{N-1}$. The process of centring can be seen as a projection on the hyperplane $\mathcal{H}^{N-1}$ with the unit vector as normal vector.  Moreover, the intersection between $\mathcal{H}^{N-1}$ and $\mathcal{S}^{N-1}$ is $\mathcal{S}^{N-2}$. We further define the following isomorphism:

\begin{equation}
\begin{array}{cccccc}
\phi & : & \mathcal{H}^{N-1}   & \to & \mathbb{R}^{N-1} &\\
 & &  {\bs{h_n}}& \mapsto & \phi(\bs{h_n})= \bs{f_n} & n=1,...,N-1,\\
\end{array}
\end{equation}

where  $\{ {\bs{h_n}}\}_{n=1,...,N-1}$ is an orthogonal base of $ \mathcal{H}^{N-1}$ and $\{\bs{f_n} \}_{n=1,...,N-1}$ is the canonical base of $\mathbb{R}^{N-1}$. We define:

$$
{\bs{h_n}} =\frac{ \sum_{i=1}^{n} \bs{e_i} - (n-1) \bs{e_{n+1}}}{\|\sum_{i=1}^{n} \bs{e_i} - (n-1) \bs{e_{n+1}}\|},
$$

with $\{\bs{e_n}\}_{n=1,...,N} $ the canonical base of $\mathbb{R}^{N}$. Note that $\phi (\mathcal{S}^{N-2})=\mathcal{S}^{N-2}$, and that is why we can work in $ \mathbb{R}^{N-1}$ and then return in $ \mathbb{R}^{N}$.\\

\subsection{The AcSel algorithm}

To use the selection-boost algorithm, we need  a grouping method $gr_{c_0}$  depending on an user-provided constant $0\leqslant c_0 \leqslant 1$. This constant determines the strength of the grouping effect. The grouping method maps each variable index $1,...,P$ to an element of $\mathcal{P}(\{1,...,P\})$ (with $\mathcal{P}(S)$ is the powerset of the set $S$, $i.e.$ the set which contains all the subsets of $S$.) . In practicals terms, $gr_{c_0}(p)$ is the ensemble of all variables which are considered to be correlated to the variable $\bs{x}_p$ and $\bs{X}_{gr_{c_0}(p)}$ is the submatrix of $\bs{X}$ containing the row which indices are in $gr_{c_0}(p)$.  If $gr_{c_0}$ is looked as function which depends on $c_0$ it should have the following properties :

 \begin{itemize}
 \item $c_0 \in [0,1]$,
\item $gr_1(p)$ is the the set of all the indices of variables perfectly correlated (positively or negatively) to variable $p$,
\item $gr_0(p)$ is the set containing the indices of all the variables,
\item if $c<c'$ then $gr_{c'} \subset gr_{c}$.
\end{itemize}

   Furthermore, we need to have a selection method $\begin{array}{ccccc} select & : & \mathbb{R}^{N \times P } \times \mathbb{R}^N& \to & \{0,1\}^P \end{array}$ which maps the design matrix $\bs{X}$ and the response variable $\bs y$ to a 0-1 vector of length $P$ with $1$ at position $p$ if the method selects the variable $p$ and 0 otherwise.\\

Here, we make the assumption that a group of correlated variables are independent realizations of the same multivariate Gaussian law. As the variables are normalized with respect to the $\mathcal{L}_2$ norm, we will use the von-Mises Fisher law \cite{sra2012short} in $\mathbb{R}^{N-1}$ thanks to the isomorphism $\phi$.  The probability density function of the von Mises-Fisher distribution for the random $P$-dimensional unit vector $\bs{x}\,$ is given by:

$$
f_{P}(\bs{x};\bs\mu, \kappa)=\tilde{K}_{P}(\kappa)\exp \left( {\kappa \bs \mu' \bs{x} } \right)
$$

where $\kappa \geqslant 0$, $ \bs\mu =(\mu_1,...,\mu_P)'$,  $\left \Vert \bs\mu \right \Vert_2 =1 \,$ and the normalization constant $C_{P}(\kappa)\,$ is equal to

$$
\tilde{K}_{P}(\kappa)=\frac {\kappa^{P/2-1}} {(2\pi)^{P/2}I_{P/2-1}(\kappa)},
$$

where $ I_{v}$ denotes the modified Bessel function of the first kind and order $v$ \cite{abramowitz1972handbook}. \\

We then use the von-Mises Fisher law to create replacement of the original variables by some simulations (see Algorithm 1) to create $B$ new design matrices $\bs{X}^{(1)},...,\bs{X}^{(B)}$.  The AcSel algorithm then applies the variable selection method $select$ to each of these matrices and returns a vector of length $P$ with the frequency of apparition of each variable. The frequency of apparition of variable $\bs{x}_{p.}$, noted $\zeta_p$ is assumed to be  an estimator of the probability $\mathbb{P}(\bs{x}_{p.} \in \mathcal{S} ) $ for this variable to be in $\mathcal{S}$. Nevertheless, both the grouping method and the choice of $c_0$  are crucial. When this constant is too small, the model is not   enough perturbed. On the other hand, when this constant is too large, variables are chosen at random. \\

\begin{algorithm}
\caption{Pseudo-code for the AcSel algorithm with $c_0$ fixed}
\begin{algorithmic}
\REQUIRE $gr_{c_0},{select}, B,c_0,P$
\STATE $\bs\zeta \leftarrow\bs{0}_P$
\FOR{$b=1,...,B$}
\STATE $\bs{X}^{(b)} \leftarrow\bs{X}$
\FOR{$p=1,...,P$}
\STATE $\bs{x}^{(b)}_{p.} \leftarrow \phi^{-1} \left(\text{random-vMF}\left(\hat{\bs \mu}(\phi(\bs{X}_{gr_{c_0}(p)} )),\hat\kappa(\phi(\bs{X}_{gr_{c_0}(p)} )\right) \right)$
\ENDFOR
\STATE $\bs\zeta \leftarrow \bs\zeta + {select}(\bs{X^{(b)}},\bs y)$
\ENDFOR
\STATE $\bs\zeta \leftarrow \bs\zeta /B$
\end{algorithmic}
\end{algorithm}

The AcSel algorithm returns the vector $\bs\zeta = (\zeta_1,...,\zeta_P)'$. One has now to choose a threshold to determine which variables are selected. In this article, we choose to select a variable $p$ if $\zeta_p=1$. In some applications, lower choices of threshold may be chosen. 

\subsection{Choosing  $c_0$}

As  we will show in the next session, the smaller the $c_0$ parameter, the higher the precision of the resulting selected variables. On the other hand, it is obvious that the probability of choosing none of the variables (\textit{i.e.} resulting in the choice of the empty set) increases as the parameter $c_0$ decreases. In the perspective of experimental planning, the choice of $c_0$ should result of a compromise between  precision and  proportion of empty models. Nevertheless, the $c_0$ parameter can be used to  introduce a confidence indicator $\gamma_p$ related to the variable $\bs{x}_{p.}$:

\begin{equation}
\gamma_p = 1 - \min_{\bs{x}_{p.} \in \hat{\mathcal{S}}_{c_0}} c_0.
\end{equation}

When $\gamma_p $ is near to zero, we know that a little perturbation is enough for variable $p$ not being in the set of selected variables. On the other hand, when $\gamma_p$ is near to one, we know that a really strong perturbation is needed for variable $p$ not being in the set of selected variables. Notice that using the confidence indicator implies using the AcSel method with all $c_0 \in [0,1]$, or at least with a discretized grid of $[0,1]$.

\subsection{Choosing the grouping method}

The simplest method for the grouping function $gr_{c_0}$ is the following:

\begin{equation}\label{gr1}
gr_{c_0}(p) = \Big\{ p' \in \{1,...,P\} \big|  |<\bs{x}_{p.},\bs{x}_{p'.}>| \geq c_0 \Big\} .
\end{equation}

 In other words, the correlation group of the variable $p$ is determined by variables whose correlation with $\bs{x}_{p.}$ is at least $c_0$. In the following this method will be refereed as the "naïve" grouping method. Nevertheless, the structure of correlation may further be taken into account using graph community clustering. Let $\bs C$ be the correlation matrix of matrix $\bs X$. Let define $\check{\bs C}$ as follows: 

$$
\check{c}_{ij}= \left\{\begin{array}{ccc}
|\check{c}_{ij}|&\text{if}&|\check{c}_{ij}| >c_0 \text{~~and~~} i\neq j\\
0& \text{otherwise}&
\end{array}\right. \cdot
$$

Then, we apply a community clustering algorithm on the undirected network with weighted adjacency matrix defined by $\check{\bs C}$.

\subsection{Conclusion}

In this section we introduced the AcSel algorithm. As reported, this method is dependant on three elements which are the initial selection function $select$, the grouping method $gr_{c_0}$, and  the constant $c_0$. Whereas any of the variable selection functions reviewed in the introduction can be used for the selection function, we provide two specific grouping methods. Furthermore, the next section will show that the choice of $c_0$ should be made in respect to practical considerations.

\section{Numerical studies}

\subsection{Introduction}

To access the performances of the AcSel algorithm, we performed numerical studies. As stated before, the AcSel algorithm can be applied to any existing variable selection method. Here, we decided to use the Lasso and forward stepwise selection. The performance of the Lasso is known to be strongly dependant on the choice of the penalty parameter $\lambda$. In our simulations, we used four criteria to choose this penalty parameter: BIC, modified BIC (BIC2) in which the estimation of the residual variance is calculated with the model including two variables,  AICc which is known to be asymptotically equivalent to cross-validation and  generalized cross-validation (GCV).\\

To demonstrate the performance of the AcSel method, we compared our method with stability selection and with a naive version of our algorithm, naiveAcSel. The naiveAcSel algorithm works as follows: estimate $\bs{\beta}$ with any variable selection method then if $gr_{c_o}(p)$, as defined in equation (\ref{gr1}) for example, is not reduced to $\{p\}$, shrink $\hat\beta_p$ to 0. The naiveAcSel algorithm is similar to the AcSel algorithm, except that it does not take into account the error which is made choosing at random a variable among a set of correlated variables.\\

We explored four situations. Let $P$ be the number of variables and $N$ the number of observations. Data are generated from model in equation (\ref{model}), assuming that $\epsilon_i \sim \mathcal{N}(0,\sigma^2)$. The variance $\sigma^2$ is chosen to reach a signal to noise ratio of 5. Exception made of situation 4, variables are simulated following a multivariate Gaussian law, with variance-covariance matrix $\Sigma$. The diagonal elements of $\Sigma$ are always set to 1. Each situation is repeated 200 times.  \\

\noindent
\textbf{Situation~1} We are in the case where $P=N=10$ and $\bs{\beta}=(1,1,1,0,0,...,0)'$. We set $\Sigma_{ij}=0$ for $ 1\leq i \neq j \leq 9$ and $\Sigma_{1,10}=0$. \\

\noindent
\textbf{Situation~2} We are in the case where $P=50$ and $N=20$ and $\bs{\beta}=(1,1,1,1 ,1,0,0,...,0)'$. We set $\Sigma_{ij}=0.5$ for $ 1\leq i \neq j \leq 50$. \\

\noindent
\textbf{Situation~3} We are in the case where $P=500$ and $N=25$ and $\bs{\beta}=(1,1,1,1 ,1,0,0,...,0)'$. We set $\Sigma_{ij}=0.5$ for $ 1\leq i \neq j \leq 500$. \\

\noindent
\textbf{Situation~4} In this situation we use gene expression from a microarray data experiment in which $N=24$. We first select the 1300 genes that were differentially expressed (stimulated versus  unstimulated).   For each repetition, we randomly select 100 genes among the 1300 and use the model in equation ($\ref{model}$) to generate the response variable.  We set  $\bs{\beta}=(1,1,1,1 ,1,0,0,...,0)'$. \\

We use 4 indicators to evaluate the abilities of our method on simulated data. We define: 

\begin{itemize}
\item recall as the ratio of the number of correctly identified variables (\textit{i.e.} $\hat\beta_i \neq 0$ and $\beta_i \neq 0$) over the number   of variables that should have been discovered  (\textit{i.e.} $\beta_i \neq 0$).
\item precision as the ratio of correctly identified variables (\textit{i.e.} $\hat\beta_i \neq 0$ and $\beta_i \neq 0$) over the number of identified variables (\textit{i.e.} $\hat\beta_i \neq 0$).
\item Fscore as the following ratio: 
$$
2 \times \frac{\text{recall}\times \text{precision}}{\text{recall}+\text{precision} } \cdot
$$
\item emptiness as the proportion of empty models (no variable is  selected) 
\end{itemize} 

Recall, precision, and Fscore are calculated over all models that are not empty. Note that our interest is focused on precision, as our goal is to select reliable variables. When $c_0=1$ the AcSel algorithm has no difference with the initially selected method $select$. When $c_0$ is decreasing toward zero we expect a profit in precision and a decrease of recall. We also calculate the Fscore which  combines both recall and precision. As an improvement of precision comes with an increase of the proportion of empty models, the best method is one with the highest precision for a given level of emptiness. \\

\subsection{Results of the simulation}

\begin{figure}[!h]
\center
\includegraphics[width=\textwidth]{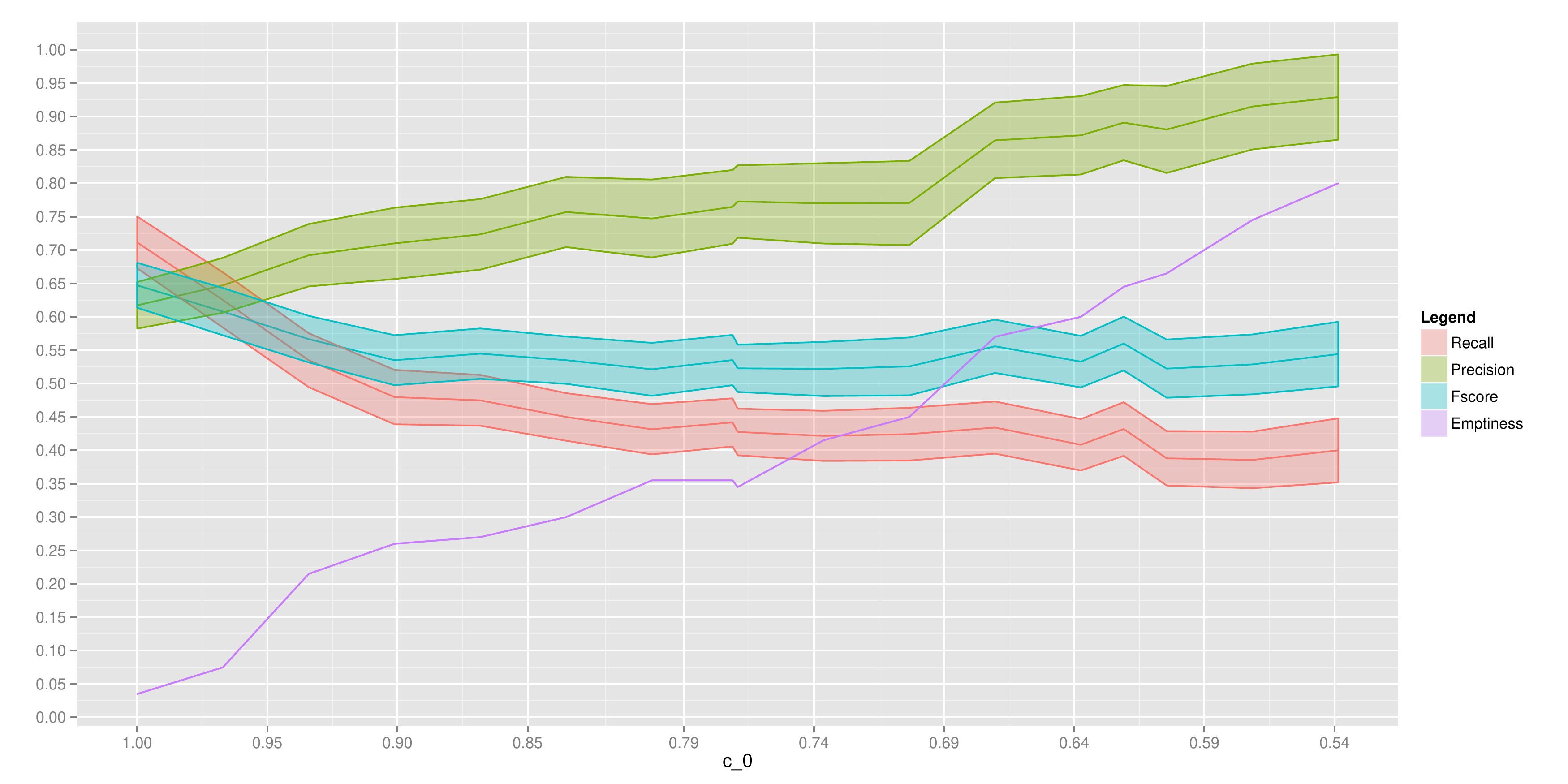}
\includegraphics[width=\textwidth]{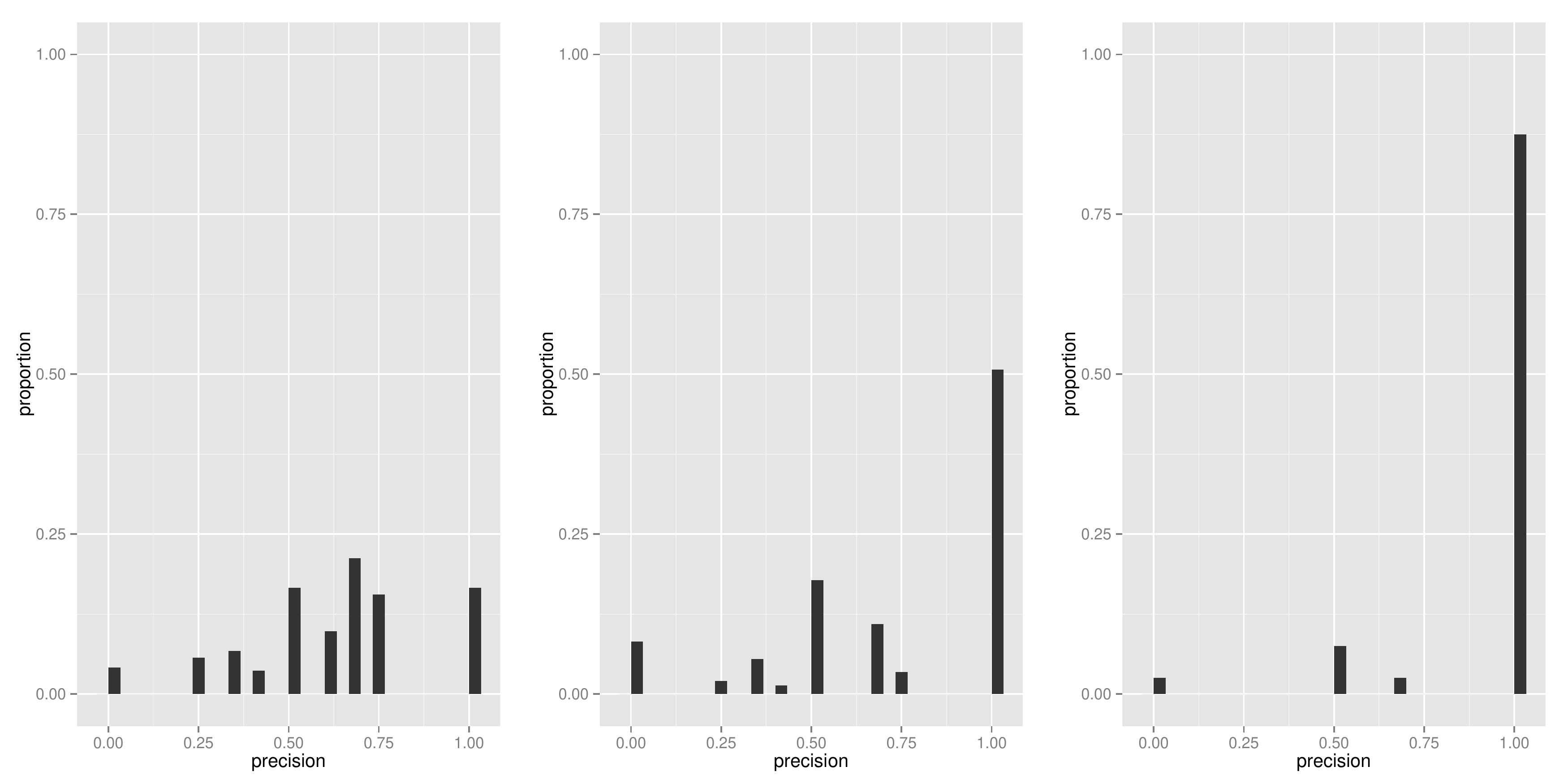}
\caption{Example of result, here Situation 1 with the Lasso with the modified BIC criterion. Top figure: evolution of the four indicators (recall, precision, Fscore and emptiness) with 95\% bandwidth confidence interval in function of $c_0$. Bottom: the distribution of the precision among all non-empty models for the highest, an intermediate, and the lowest $c_0$.}
\label{ff}
\end{figure}

Only an extract of the results is presented in the main part of the article; full results are available in supporting informations. We first analyse the results for each pair of selection method and situation. We show the evolution of the four criteria (precision, recall, Fscore and emptiness) in function of the decrease of $c_0$. When $c_0=1$, the AcSel algorithm is equivalent to the initial  variable selection method. As our main focus is on precision, we add three histograms representing the evolution of the precision distribution for the highest, an intermediate and the lowest $c_0$. Figure \ref{ff} shows the result for the Lasso with the modified BIC  in Situation 1 (other Situations for BIC2 can be found in supplementary figures \ref{st1}, \ref{st2}, \ref{st3} and \ref{st4}) . In this example, we succeed to improve  precision from 0.63 to 0.93. Ohter variable selection methods show interesting improvment of precision: the gain in precision in the lowest for the Lasso with the BIC criterion. This is not suprising since this method reaches the highest level of precision when $c_0=1$. On the other hand, the Lasso with AICc or GCV (see Figures \ref{ci1}, \ref{ci2}, \ref{ci3}, \ref{ci4}) present the greatest improvement in precision with the decrease of $c_0$: in Situation 2, for the Lasso with GCV, precision improves from 0.25 to 0.75. However, as shown by the histograms of the precision, the proportion of models for which precision reaches one increases with the decrease of $c_0$. The Fscore remains either stable or shows a small decrease indicating that the loose in recall is compensated by the increase of precision. In other words, our method allows to choose the desired trade-off between recall and precision (see Figures \ref{ci1}, \ref{ci2}, \ref{ci3}, \ref{ci4}).  \\

\begin{figure}[!h]
\center
\includegraphics[width=\textwidth]{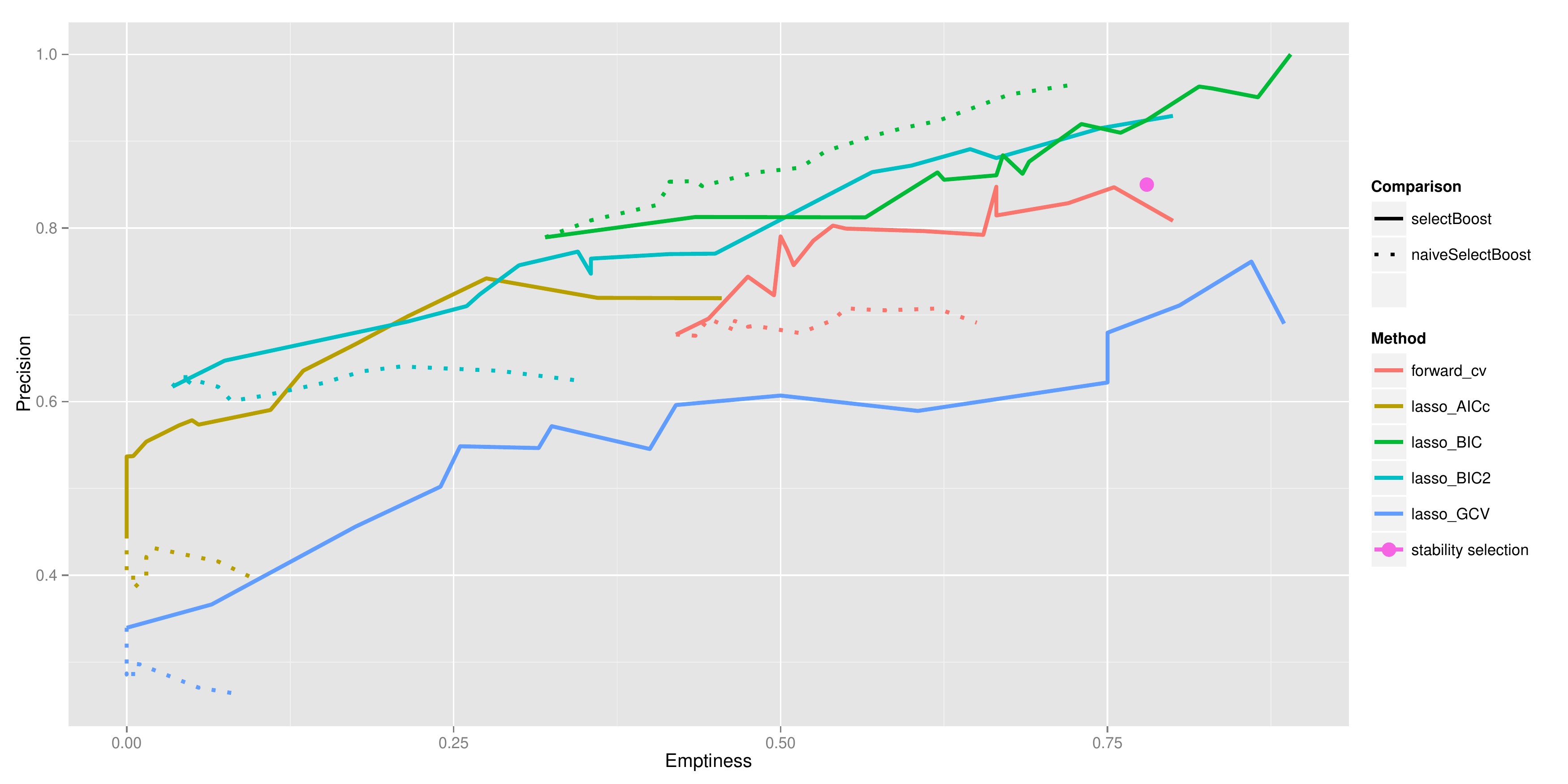}
\caption{Precision in function of emptiness for all tested method for Situation 1. The AcSel algorithm is compared to both stability selection and the naiveAcSel algorithm.}
\label{fff}
\end{figure}

\begin{figure}[!h]
\center
\includegraphics[width=\textwidth]{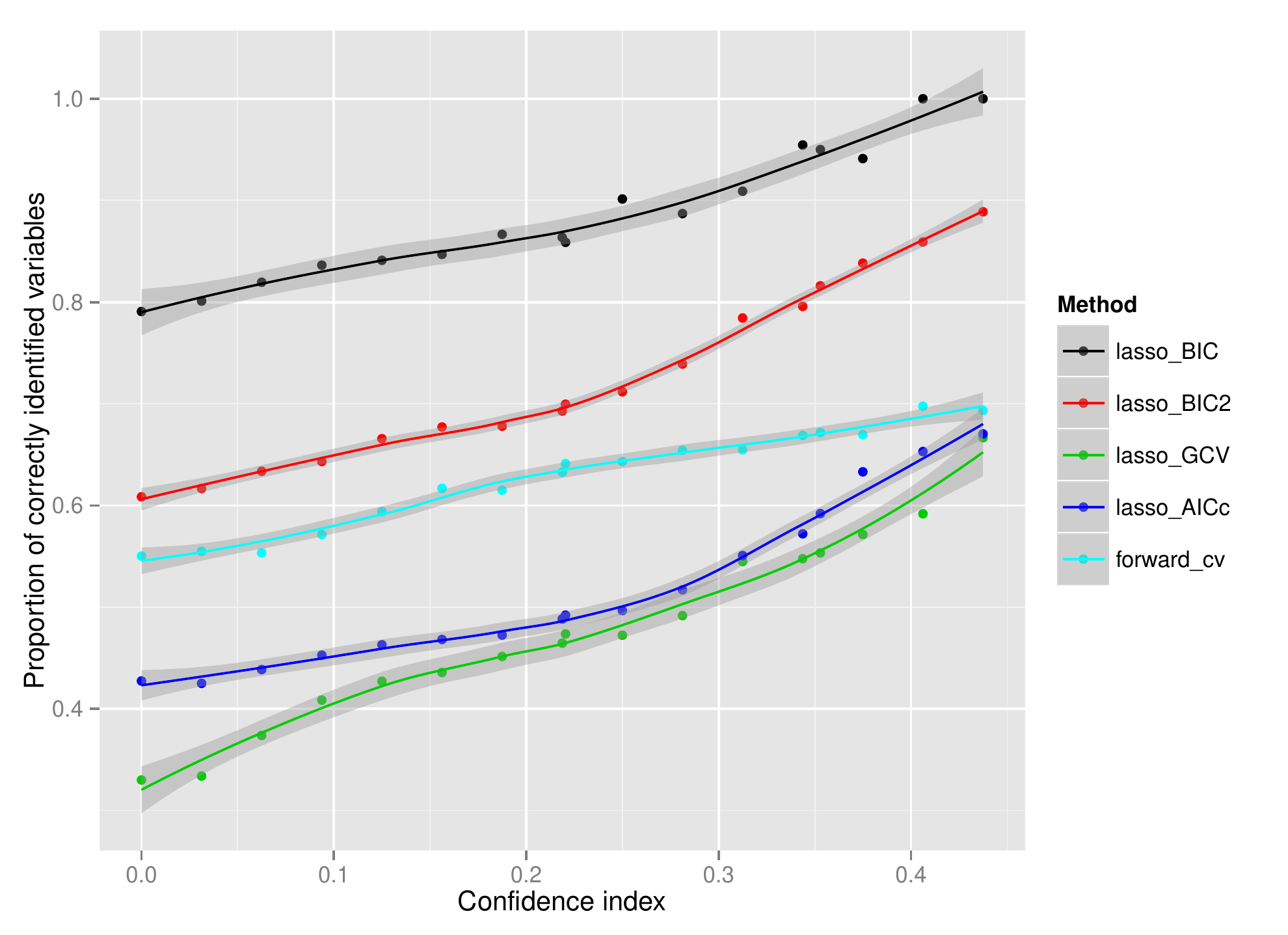}
\caption{The proportion of correctly  identified variables is plotted in function of the confidence indicator that we defined in Equation (6.8) for Situation 1. As expected, the greater the confidence indicator, the higher the proportion of correctly identified variables.}
\label{ffff}
\end{figure}

As our interest is focused on precision, our goal is to reach the highest precision with the lowest proportion of empty models. In this context, one interesting fact about the AcSel method is that the method of choice of the penalty parameter in the Lasso is no more crucial. Indeed, as shown in Figure \ref{fff}, the precision of each method is similar at a given proportion of emptiness. Nevertheless, depending of the situation, the choice of the penalty parameter by AICc (see Figures \ref{ci1}, \ref{ci2}, \ref{ci3}, \ref{ci4}) or GCV (see Figure \ref{ffff}) may lead to worse outcomes, even if there is an increase of precision with the confidence index. \\

Except in one case (the Lasso with choice of penalty parameter through the BIC criterion, see Figure \ref{fff}), the AcSel algorithm shows its superiority over the naiveAcSel algorithm. The error which is made when choosing randomly a variable among a set of  correlated variables  conduces to further wrong choice of variables. While the intensive simulation of our algorithm allows to take into account this error, the naiveAcSel does not. The superiority of the naive algorithm in Situation 1 for the Lasso with BIC criterion may be the consequence of the small size of the data and the low correlation setting. Other situations are shown in Figures \ref{C01}, \ref{C02}, \ref{C03} and \ref{C04}. \\

Finally we compare the AcSel algorithm with stability selection. Stability selection use a re-sampling algorithm to determine which of the variable which are included in the model are robust. In our simulation, stability selection shows  performances with relative high precision but also high proportion of empty models. Moreover, in contrast to the AcSel algorithm, stability selection does not allow to choose a convenient precision-emptiness trade-off.\\
 
In the previous section, we mentioned the possibility of using AcSel to obtain a confidence indicator, corresponding to one minus the lowest $c_0$ for which a variable is selected. For each situation, we plot the proportion of correctly identified variables in function of the confidence indicator (Figure \ref{ffff} for Situation 1 and Supplemental Figures for the others). The proportion of correctly identified variables increases with the increase of the confidence indicator.  

 \FloatBarrier
 
\section{Application to a real dataset}

We decided to apply our algorithm to the diabetes dataset used by Efron \textit{et al.} \cite{efron2004least}. This dataset contains 10 variables which are age, sex, body mass index, average blood pressure and six serum measurements and a quantitative response of interest that is a measure of the evolution of the diabetes disease one year after baseline. As proposed, we included interactions term, resulting in an 64 explanatory variables dataset. \\

\begin{figure}[!h]
\center
\includegraphics[width=\textwidth]{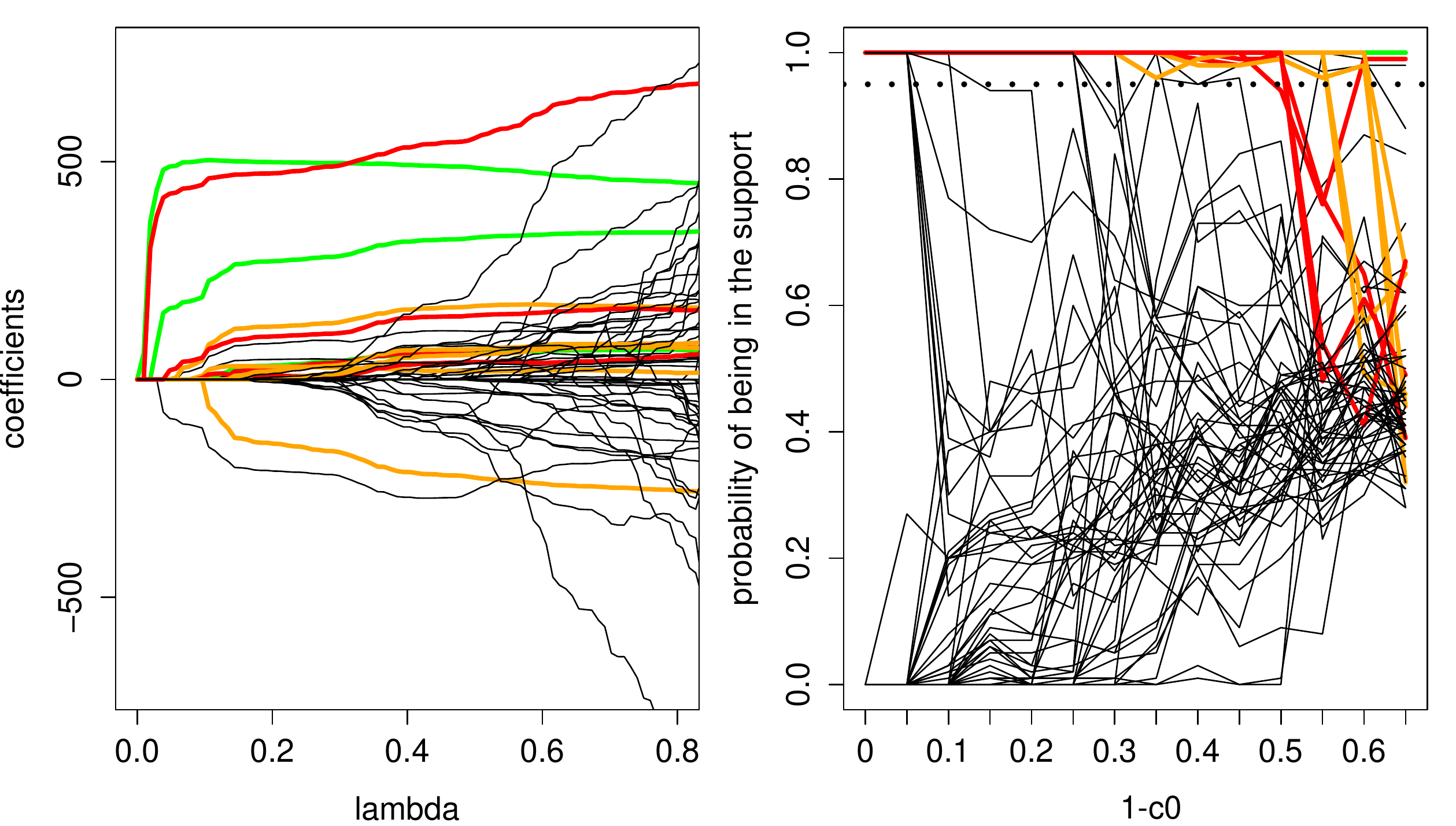}
\caption{Colors: the green is for the most reliable variables selected by the AcSel algorithm (confidence indicator of 0.65; orange is for intermediate confidence (0.55) and red for low confidence (0.45). Right: evolution of the coefficients in the lasso regression when the sparsity parameter lambda is varying. right: evolution of the probability of being in the support of the regression when the confidence indicator is varying. For the two graphics, each line maps with a variable.}
\label{diabetes}
\end{figure}

We use a wide range of the $c_0$ parameter, starting from 1 to 0.35 by step of 0.05 (see Figure \ref{diabetes} right). For each step, the probability of being included in the support is calculated with 500 simulations as described in the Algorithm 1. We set the threshold to 0.95 to avoid numerical instability. We used our algorithm with the Lasso and selected the regularization parameter with the AICc. \\

When $c_0=1$, our algorithm is equivalent to the Lasso and ends with a selection of 22 variables. At the opposite, when using the maximal $x_0$, our algorithm ends with a selection of only two variables : the body mass index and the average blood pressure. The interesting point is that the these two variables are neither the two first variables selected by the Lasso or the two variables with the highest coefficients (see Figure \ref{diabetes} left). This demonstrates that our algorithm can be very usefull to determine which are the variables that are selected with confidence.

\section{Discussion and conclusion}

We introduced the AcSel algorithm which uses intensive computation to select variables with high precision. The user  has the choice between using this algorithm to produce an confidence indicator, or choosing an appropriate precision-emptiness trade-off to select variables with confidence. The main idea of our algorithm is to take into account the correlation structure of the data and thus use intensive computation to select the reliable variables. We prove the performance of our algorithm through simulation studies. In some situations, where many regressions have to be made (in network reverse-engineering in which we have a regression per vertex) our algorithm may be used in an experimental design approach. We apply our algorithm on a real dataset and we found some interesting informations. The AcSel algorithm is a powerfull tool that can be used in every situation where reliable and robust variable selection has to be made. 

\bibliographystyle{alpha}
\bibliography{biblio}

\newpage

\Huge

\textbf{Supplementary information}

\normalsize

\vspace{2cm}

\setcounter{section}{0}
\renewcommand\thesection{\Alph{section}}
\setcounter{figure}{0}

\renewcommand\thefigure{\thesection.\arabic{figure}} 
\section{Confidence index}

For the four simulation situations, we plot the proportion of correctly identified variables against the confidence index. 

\begin{figure}[ht]
\center
\includegraphics[width=\textwidth]{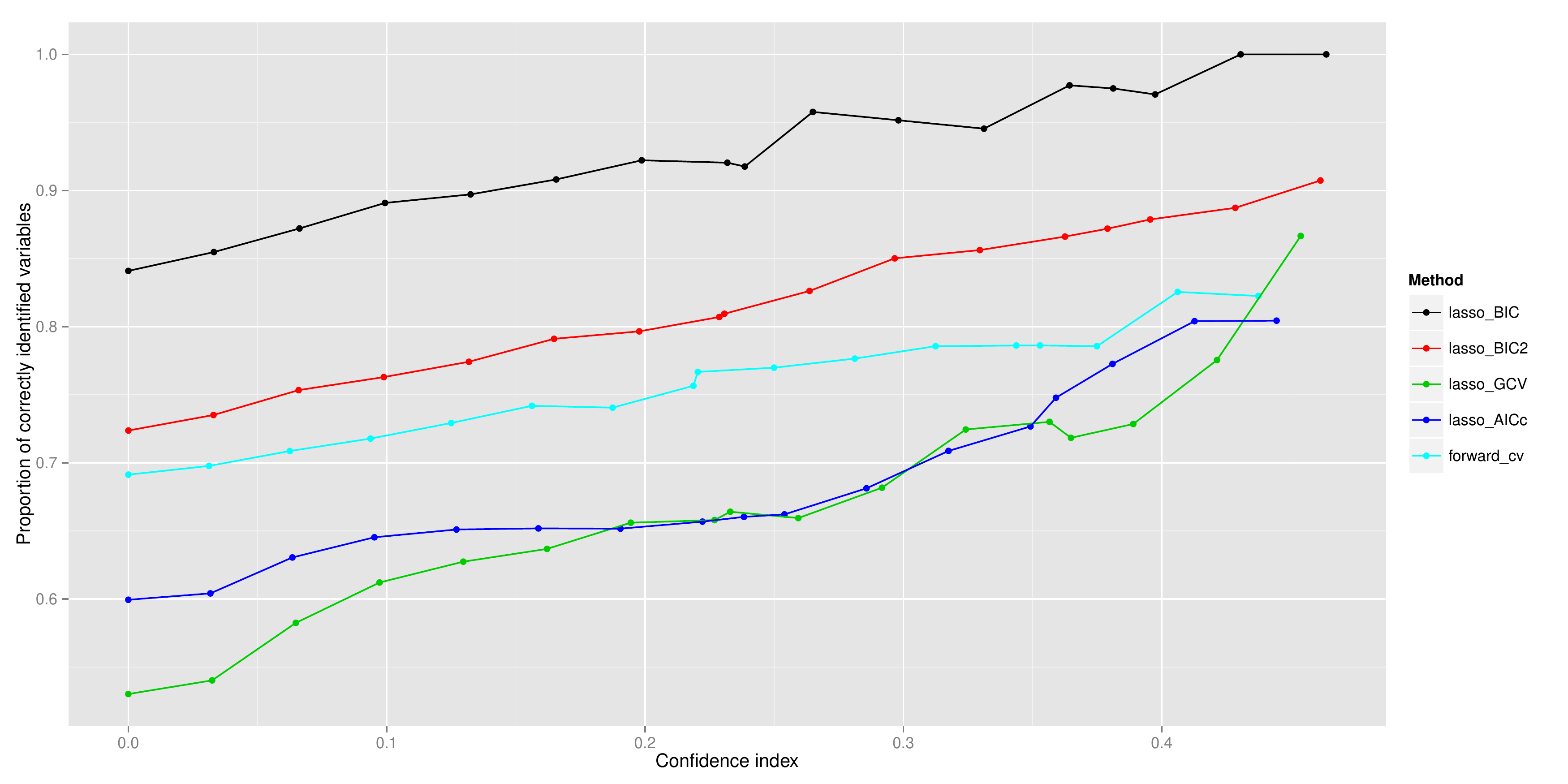}
\caption{Situation 1}
\label{ci1}
\end{figure}

\begin{figure}[ht]
\center
\includegraphics[width=\textwidth]{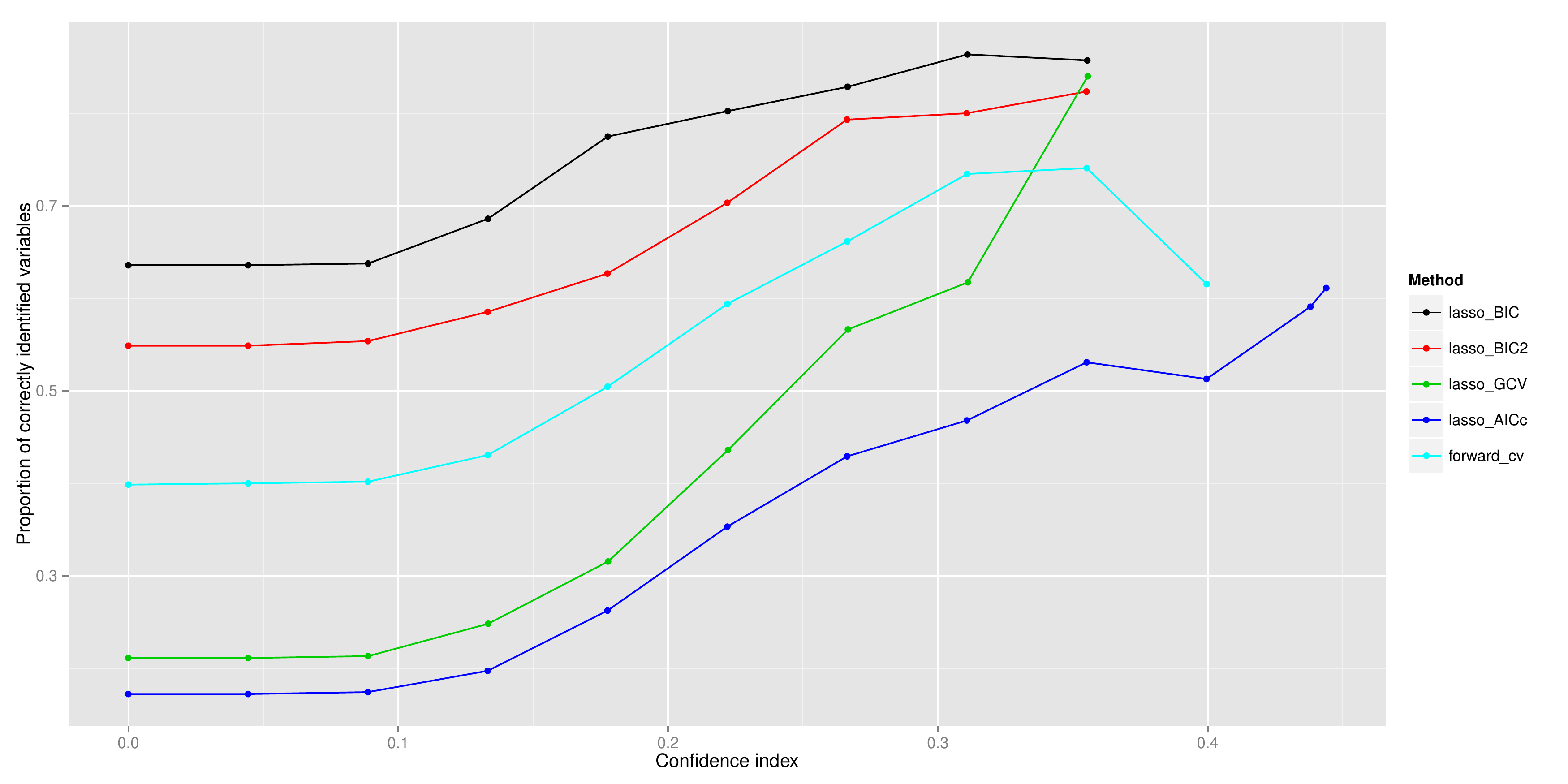}
\caption{Situation 2}
\label{ci2}
\end{figure}

\begin{figure}[ht]
\center
\includegraphics[width=\textwidth]{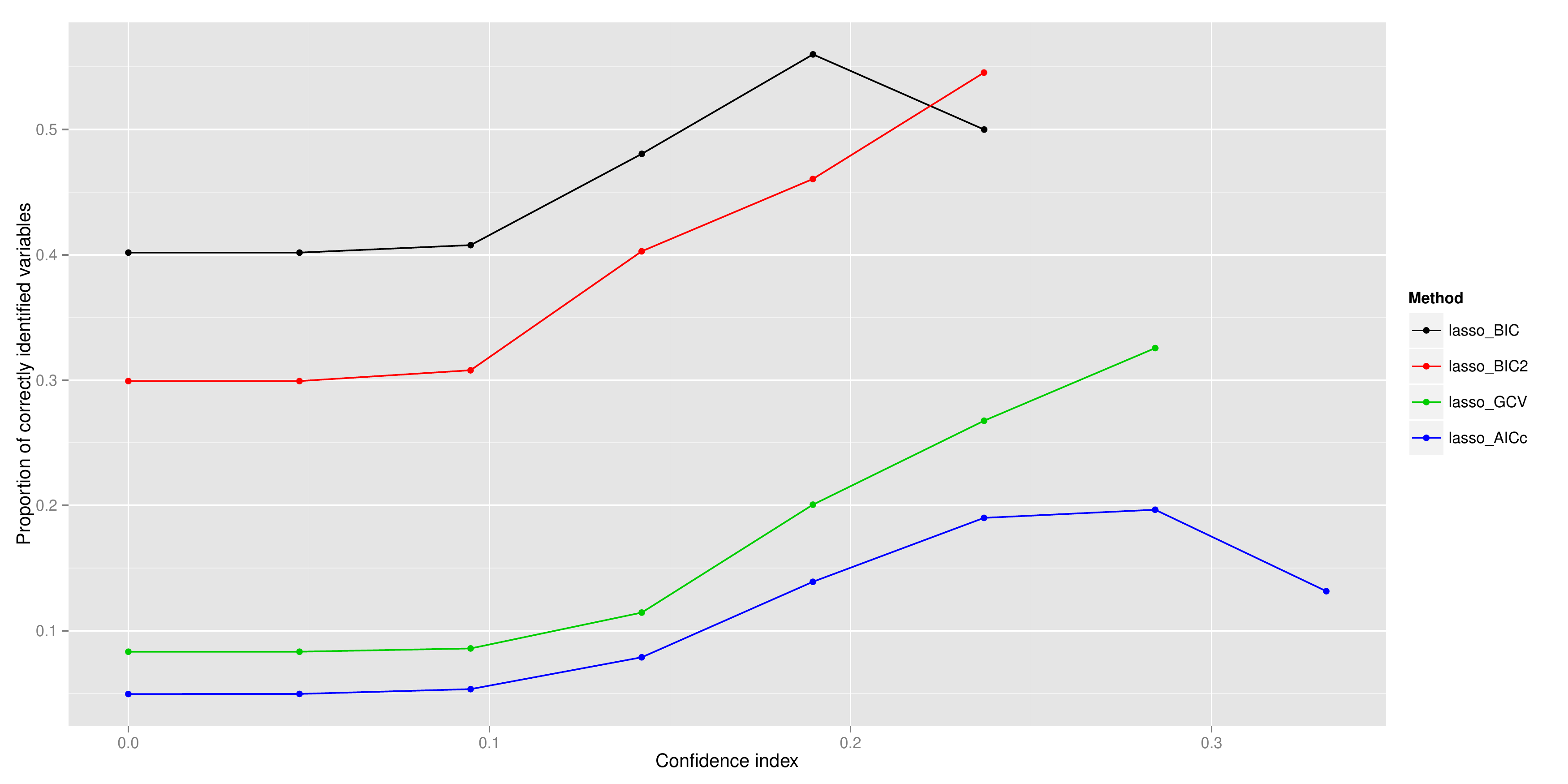}
\caption{Situation 3}
\label{ci3}
\end{figure}

\begin{figure}[ht]
\center
\includegraphics[width=\textwidth]{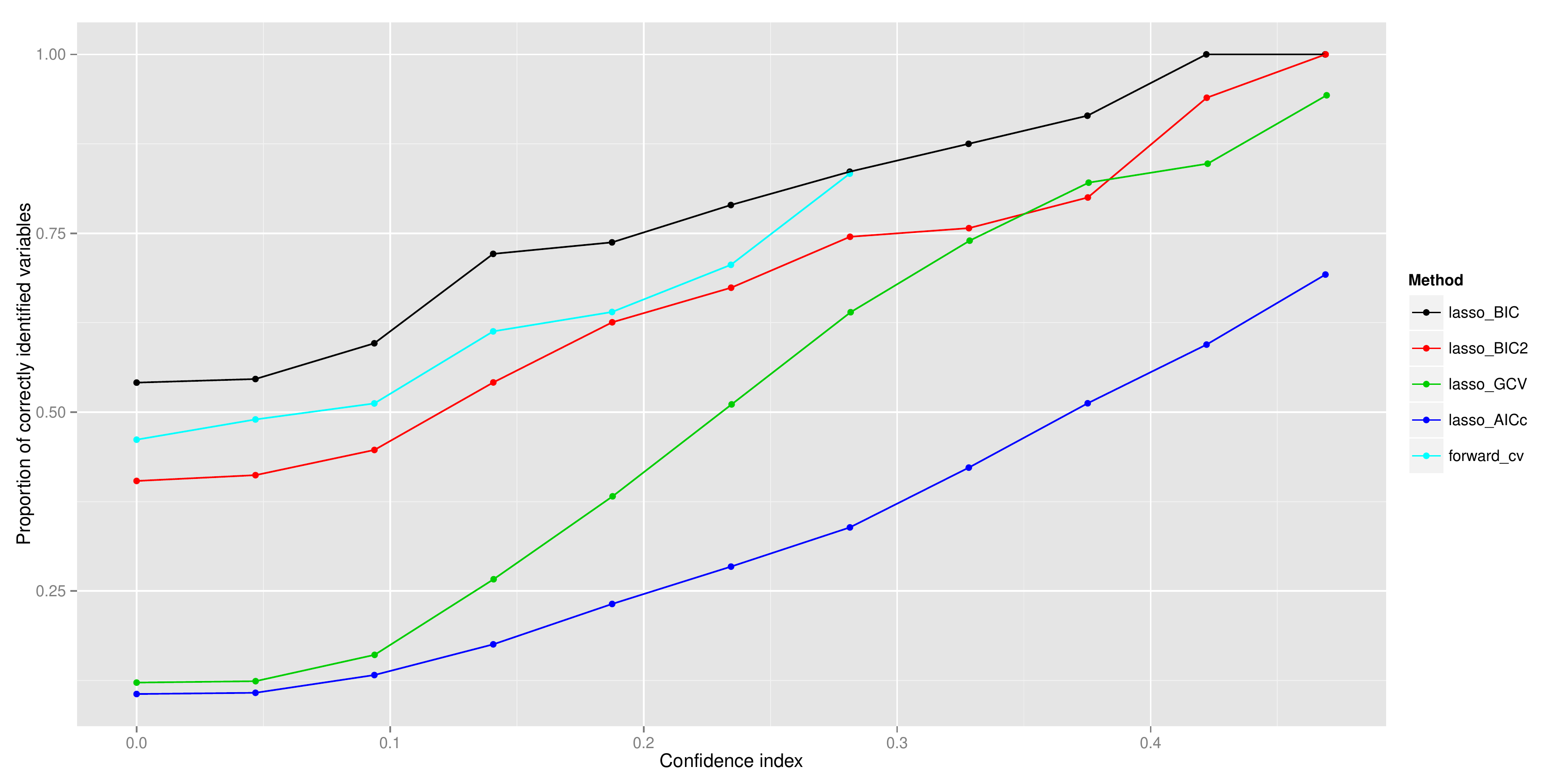}
\caption{Situation 4}
\label{ci4}
\end{figure}

\FloatBarrier
\newpage

\section{Comparisons}

\setcounter{figure}{0}

For the four simulation situations, we show the performance of our algorithm against the performance of the naive AcSel algorithm and the Stability Selection algorithm. The best algorithm is the one with the lowest proportion of empty models with the highest precision.  

\begin{figure}[ht]
\center
\includegraphics[width=\textwidth]{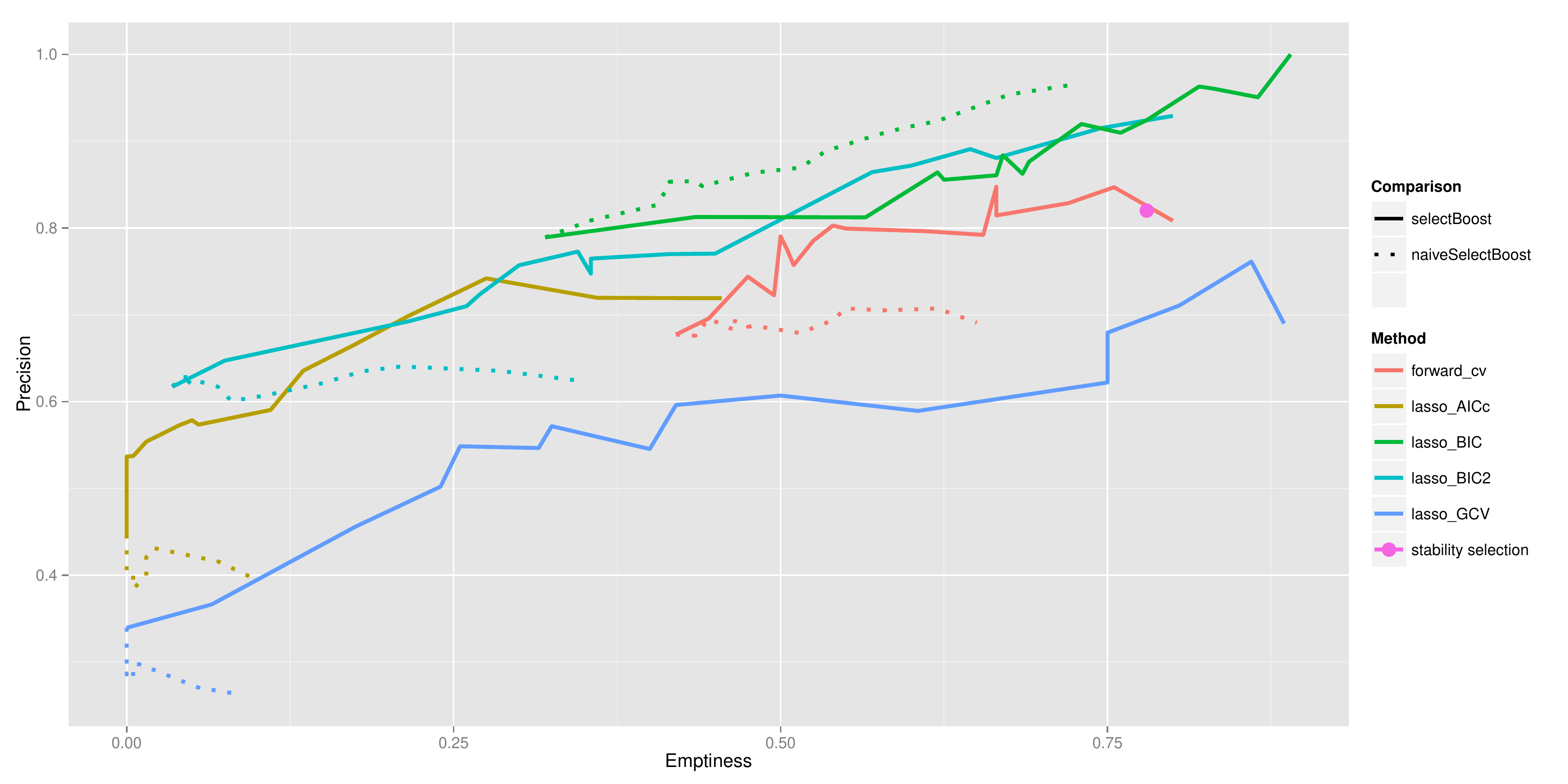}
\caption{Situation 1}
\label{c01}
\end{figure}

\begin{figure}[ht]
\center
\includegraphics[width=\textwidth]{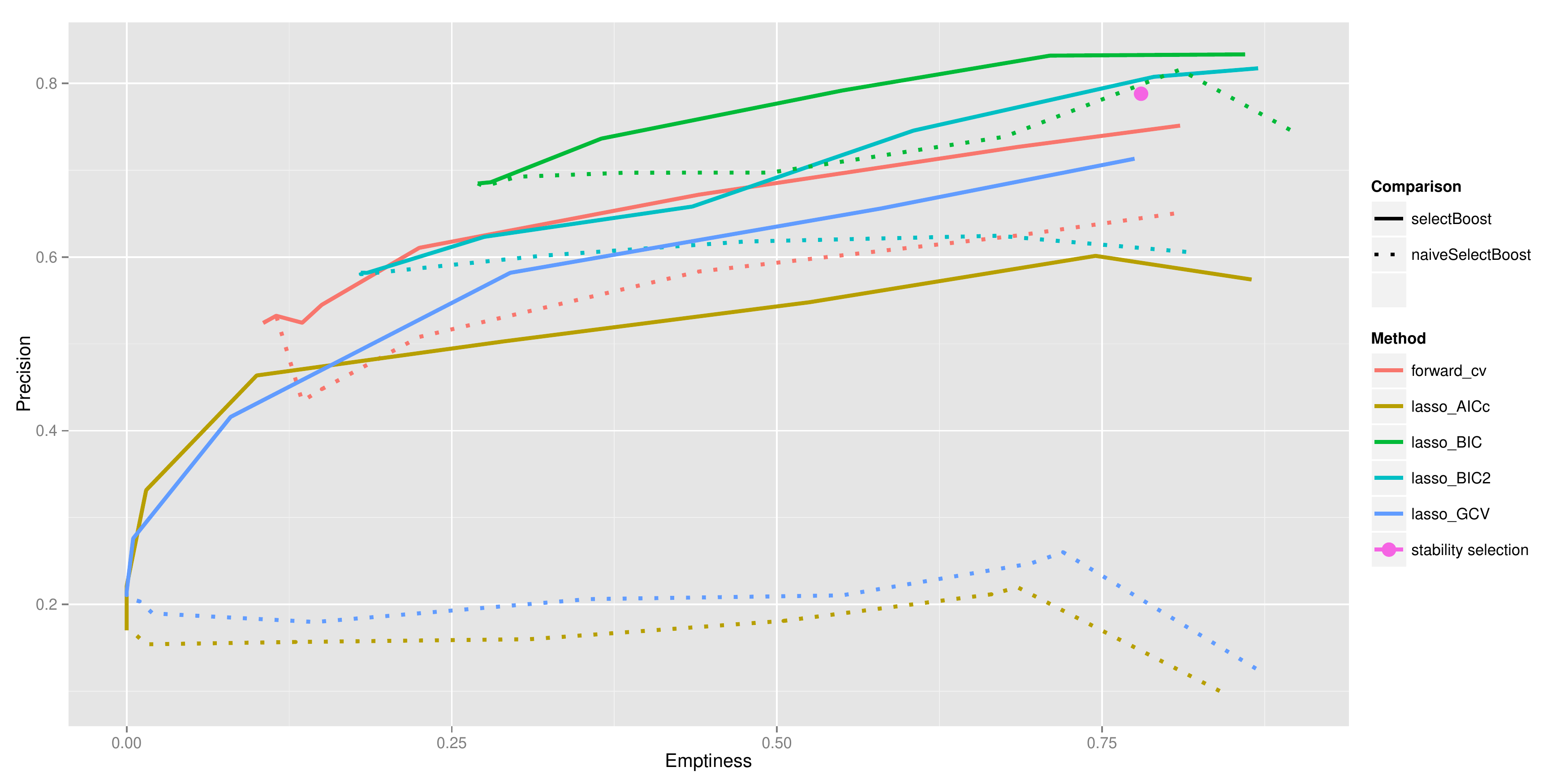}
\caption{Situation 2}
\label{c02}
\end{figure}

\begin{figure}[ht]
\center
\includegraphics[width=\textwidth]{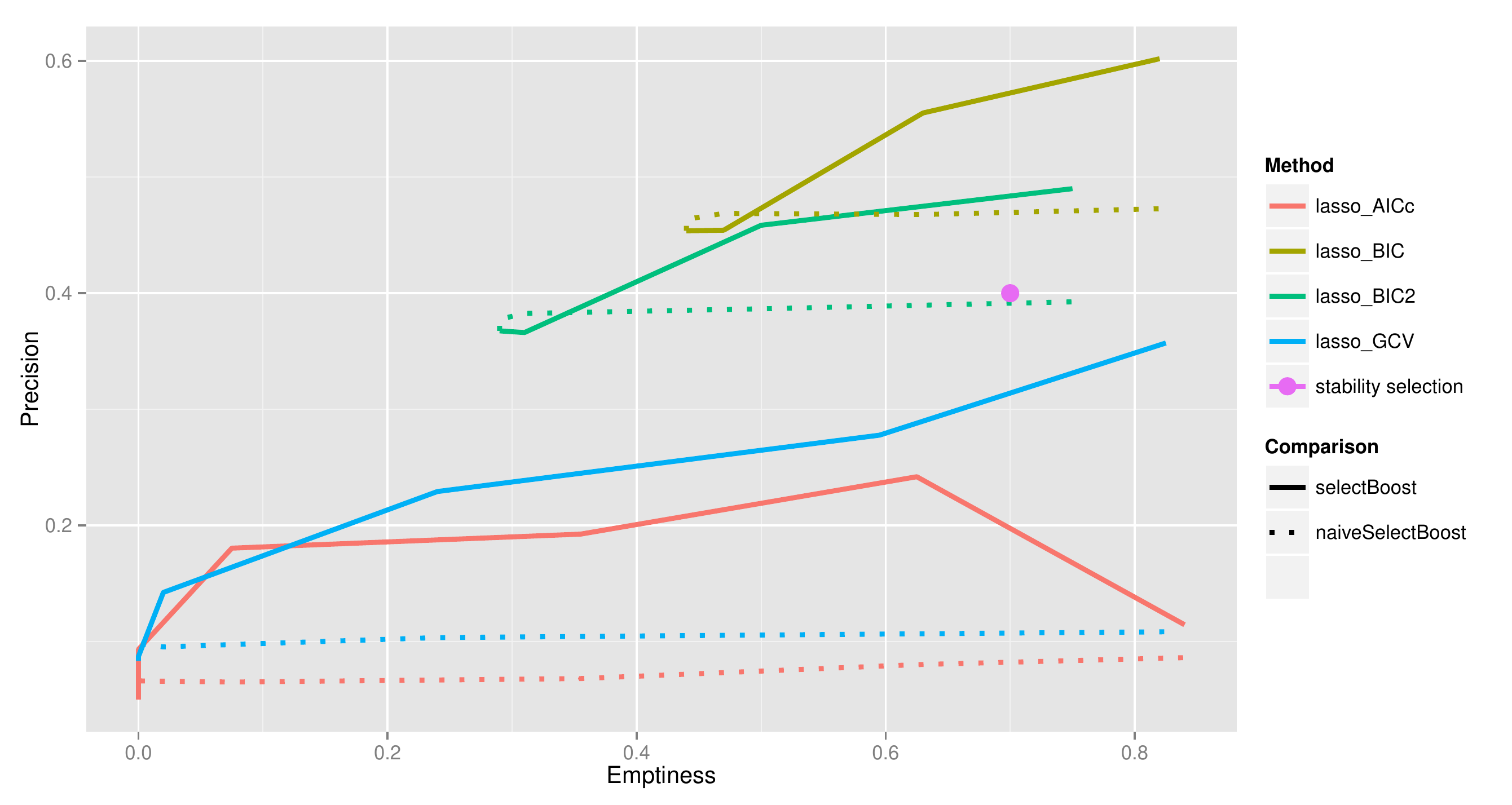}
\caption{Situation 3}
\label{c03}
\end{figure}

\begin{figure}[ht]
\center
\includegraphics[width=\textwidth]{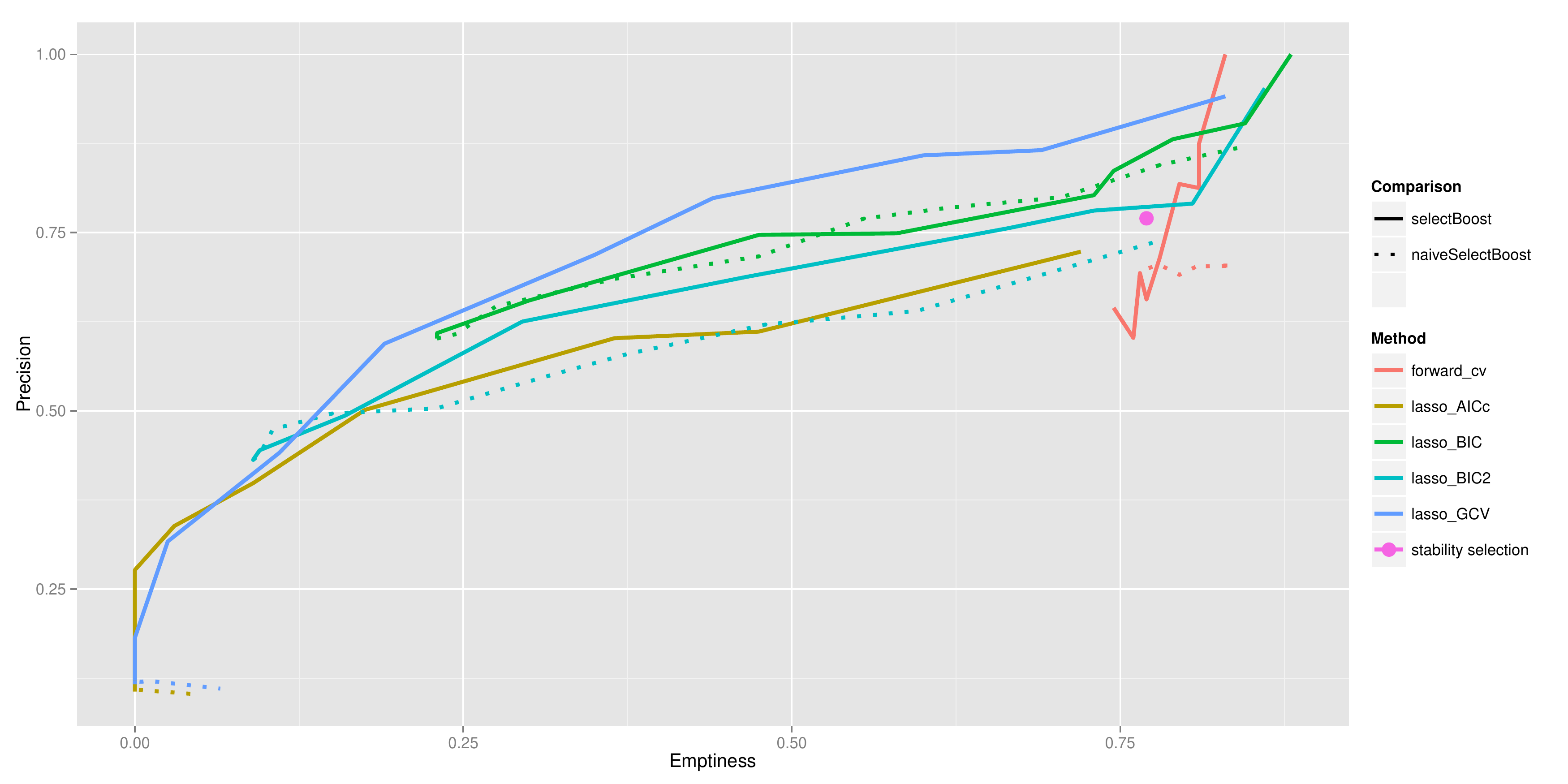}
\caption{Situation 4}
\label{c04}
\end{figure}

\FloatBarrier

\newpage

\section{Example of results: modified BIC (BIC2)}
\setcounter{figure}{0}

In  this section we show the evolution of recall, precision, Fscore and proportion of empty models with 95\% confidence interval in function of the $c_0$ parameter. We also show three histograms with the evolution of the distribution of the precision for three $c_0$ (see legends)
\begin{figure}[ht]
\center
\begin{tabular}{c}
\includegraphics[width=\textwidth]{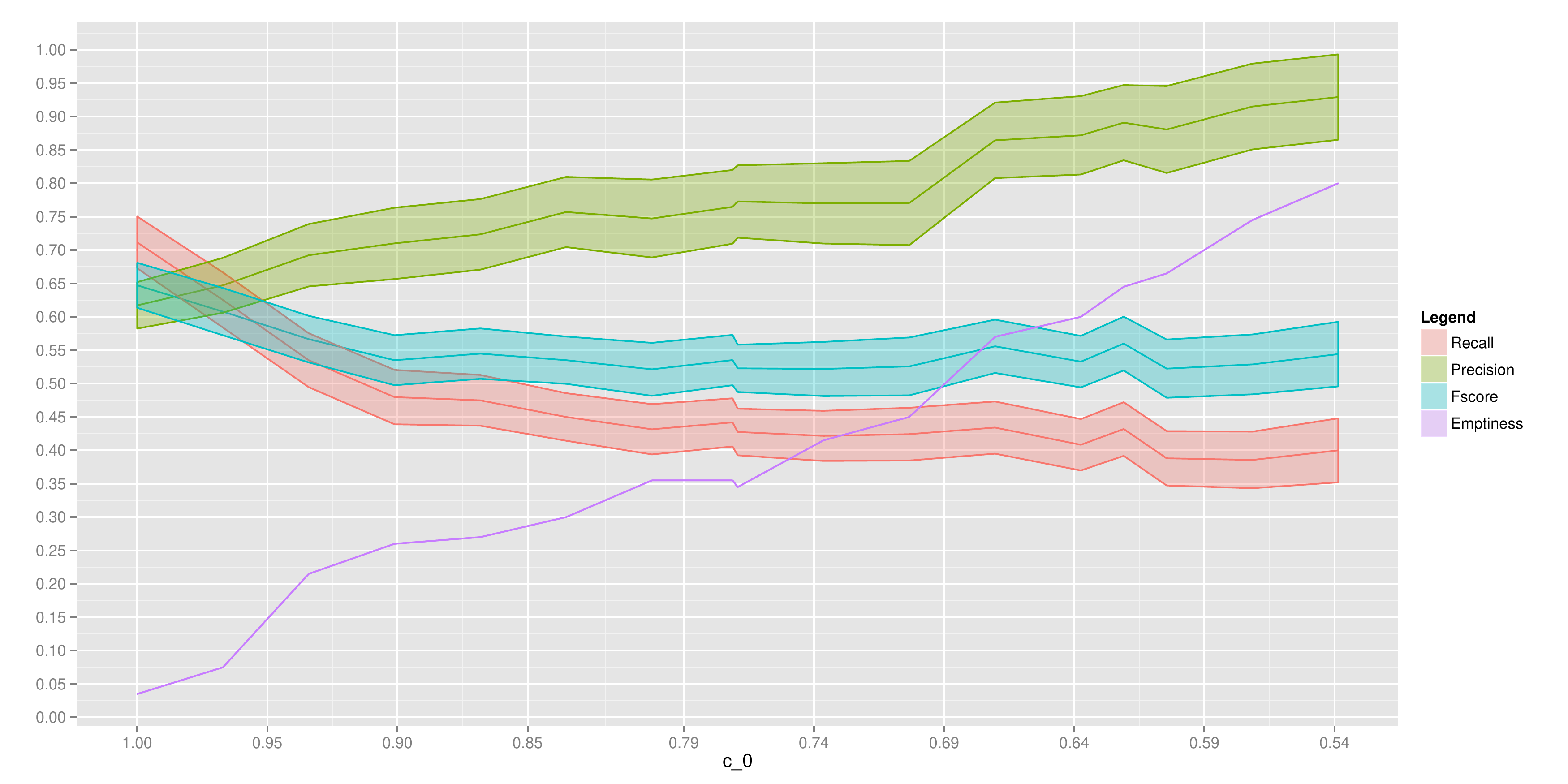}\\
\includegraphics[width=\textwidth]{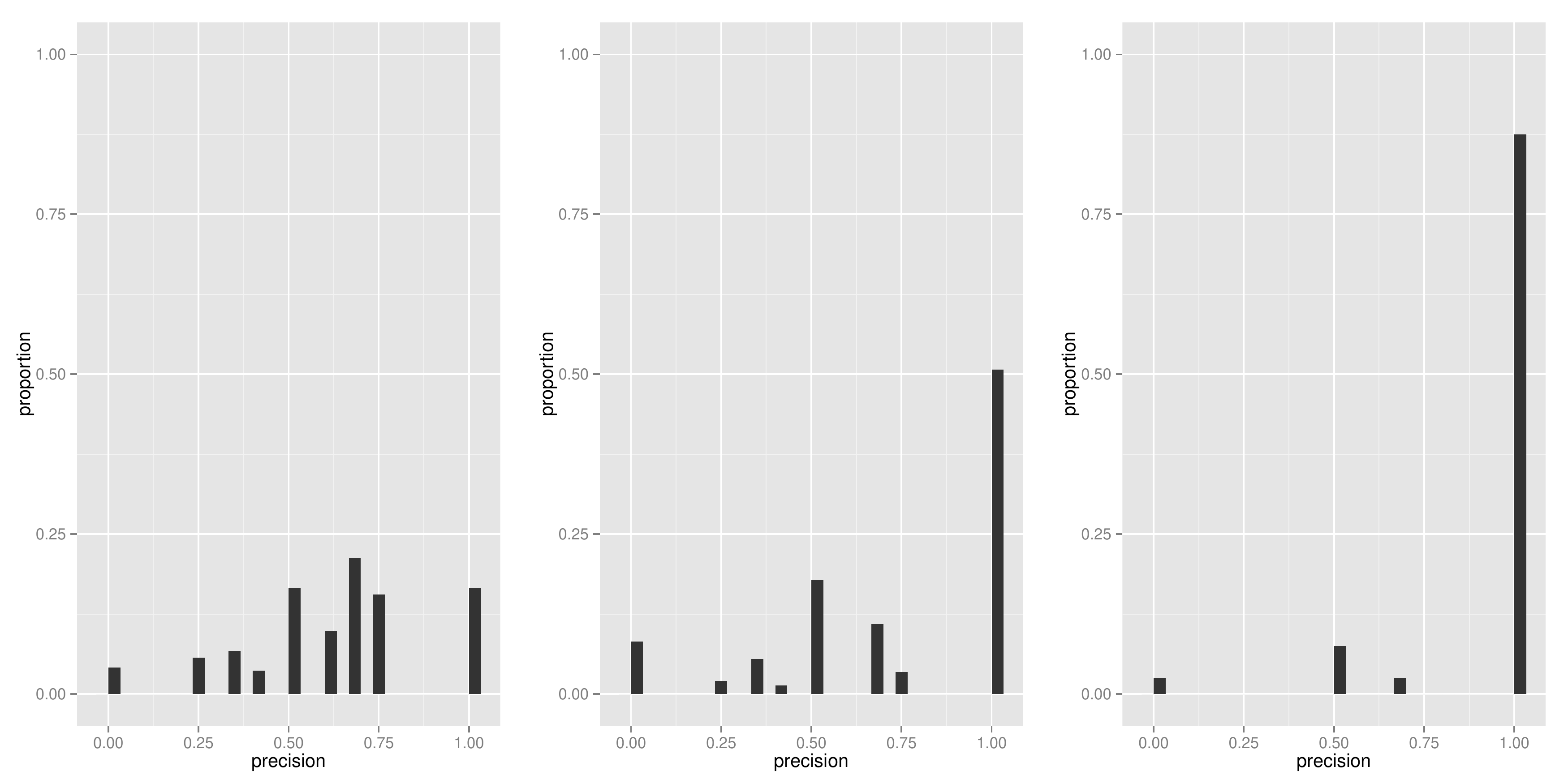}
\end{tabular}
\caption{Situation 1. The histograms show the evolution of the distribution of the precision. From left to right: $c_0=1,0.79,0.54$}
\label{st1}
\end{figure}

\begin{figure}[ht]
\center
\begin{tabular}{c}
\includegraphics[width=\textwidth]{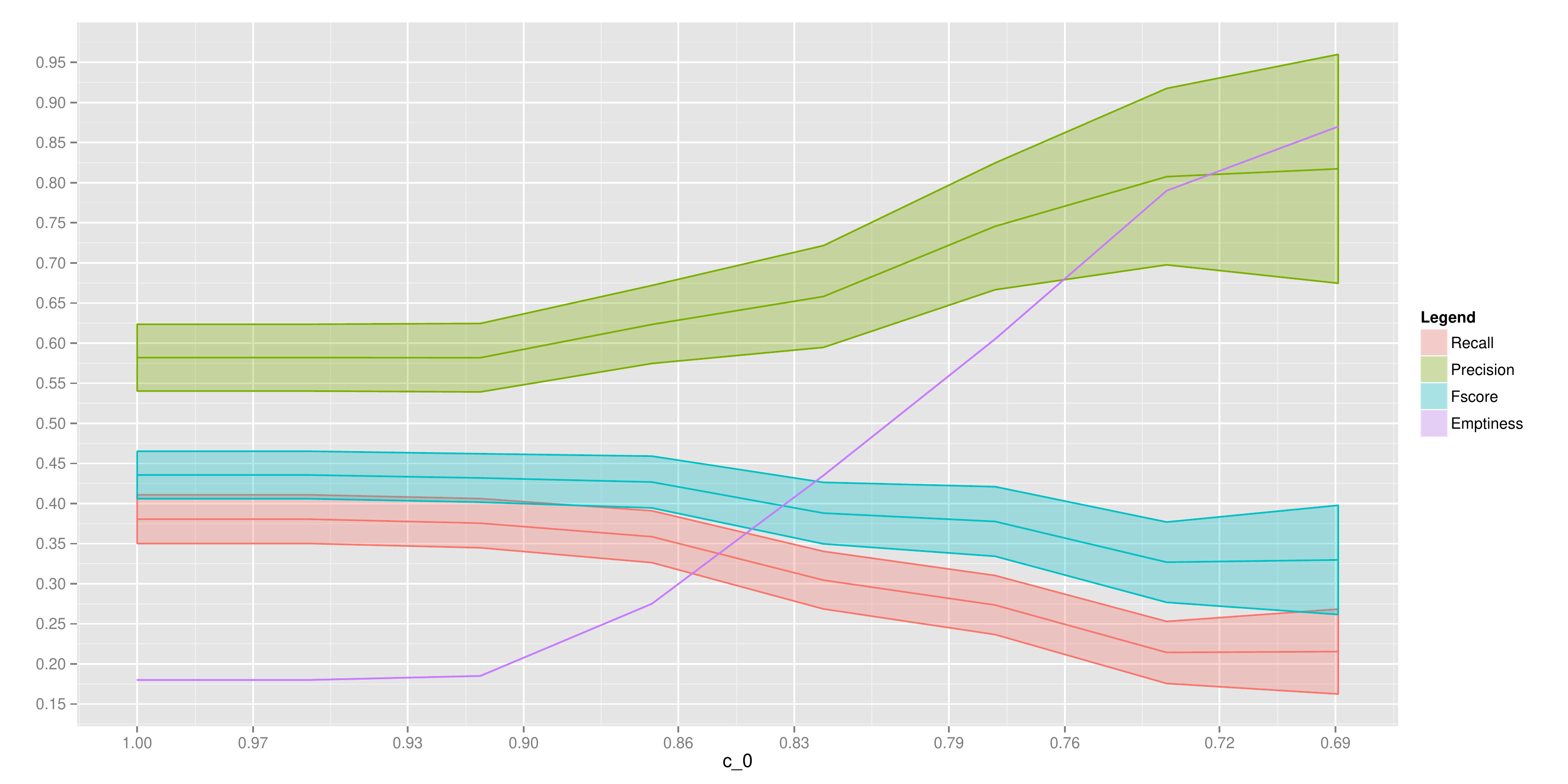}\\
\includegraphics[width=\textwidth]{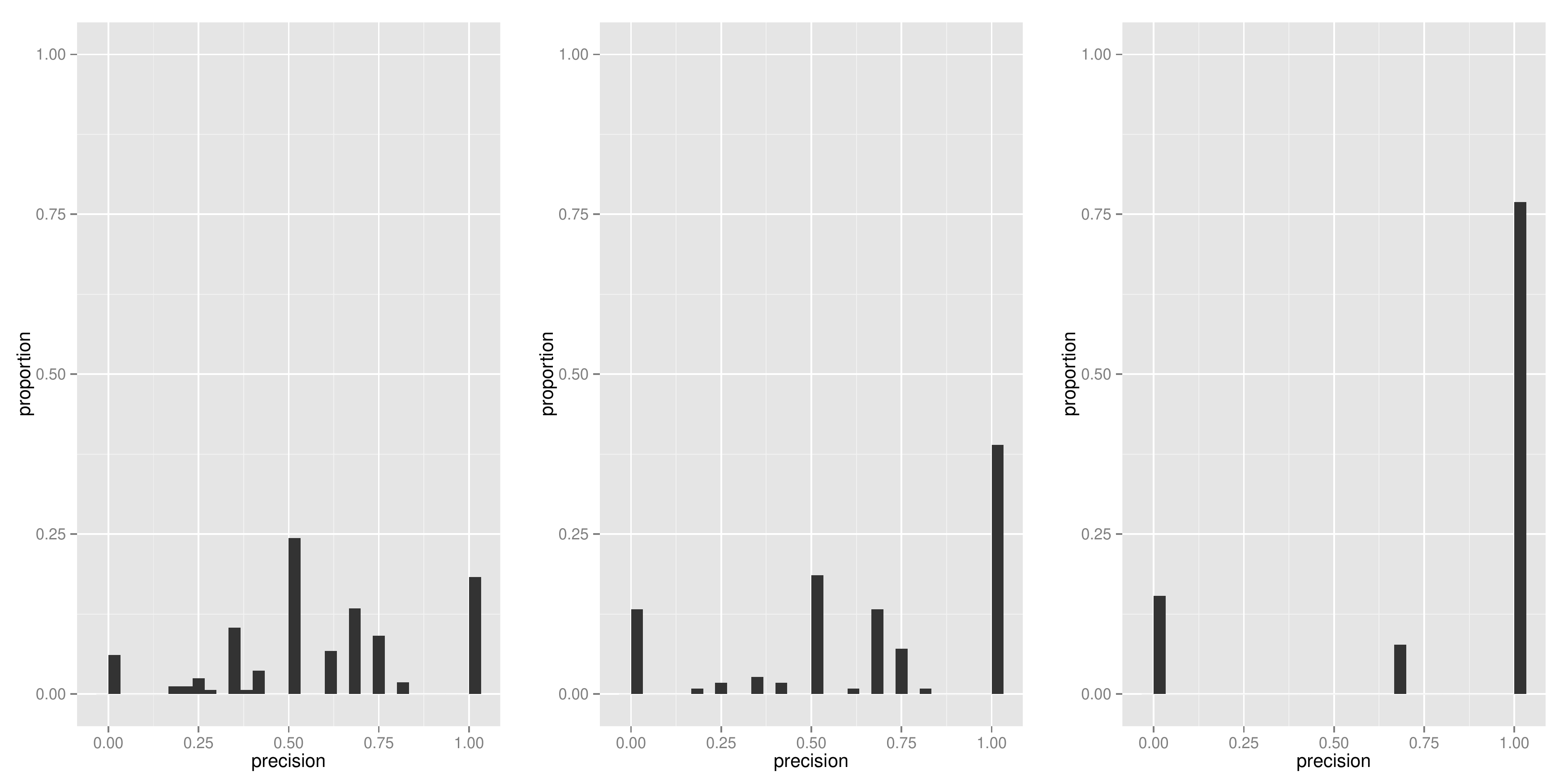}
\end{tabular}
\caption{Situation 2. The histograms show the evolution of the distribution of the precision. From left to right: $c_0=1,0.79,0.69$}
\label{st2}
\end{figure}

\begin{figure}[ht]
\center
\begin{tabular}{c}
\includegraphics[width=\textwidth]{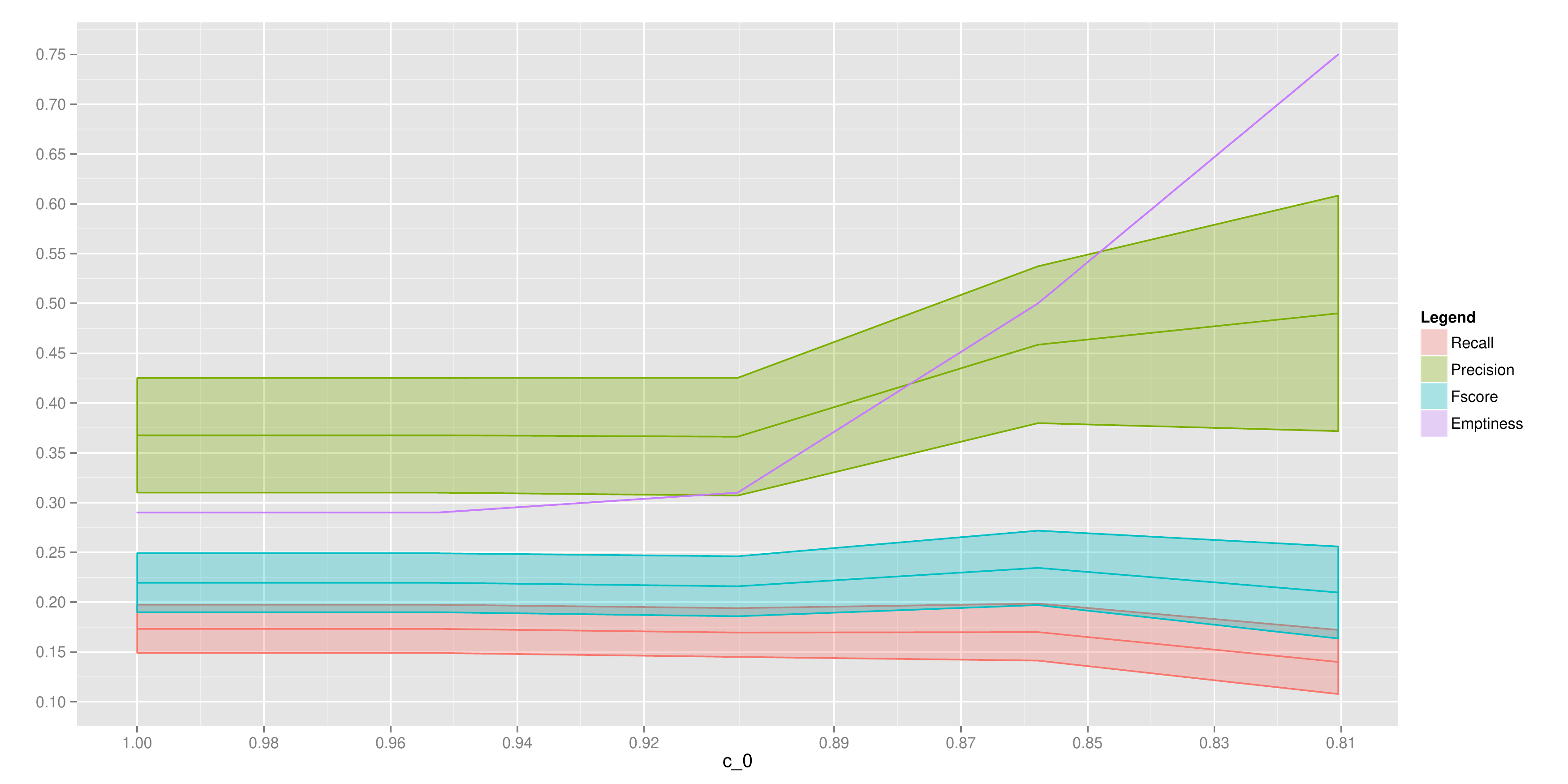}\\
\includegraphics[width=\textwidth]{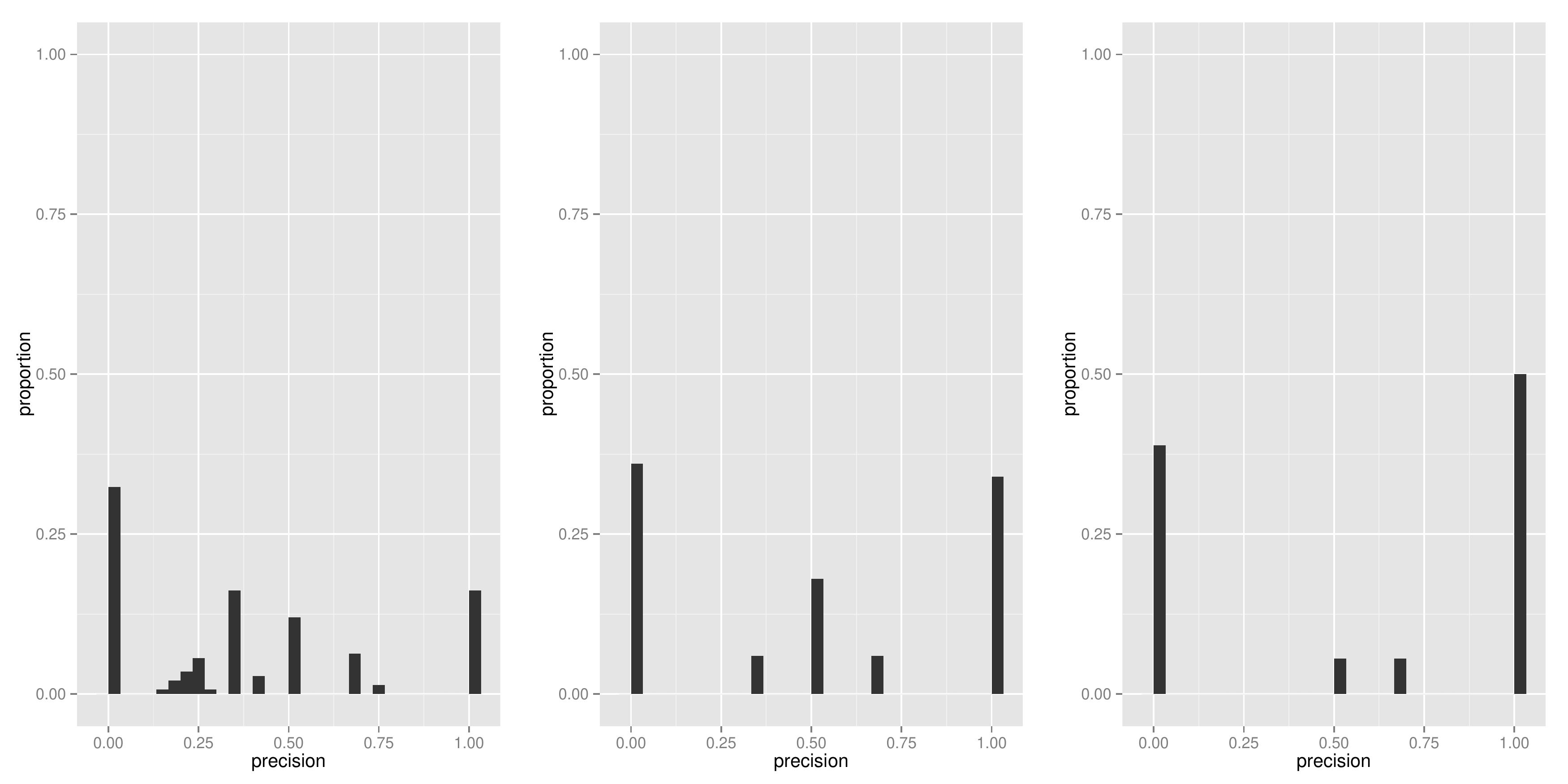}
\end{tabular}
\caption{Situation 3. The histograms show the evolution of the distribution of the precision. From left to right: $c_0=1,0.9,0.81$}
\label{st3}
\end{figure}

\begin{figure}[ht]
\center
\begin{tabular}{c}
\includegraphics[width=\textwidth]{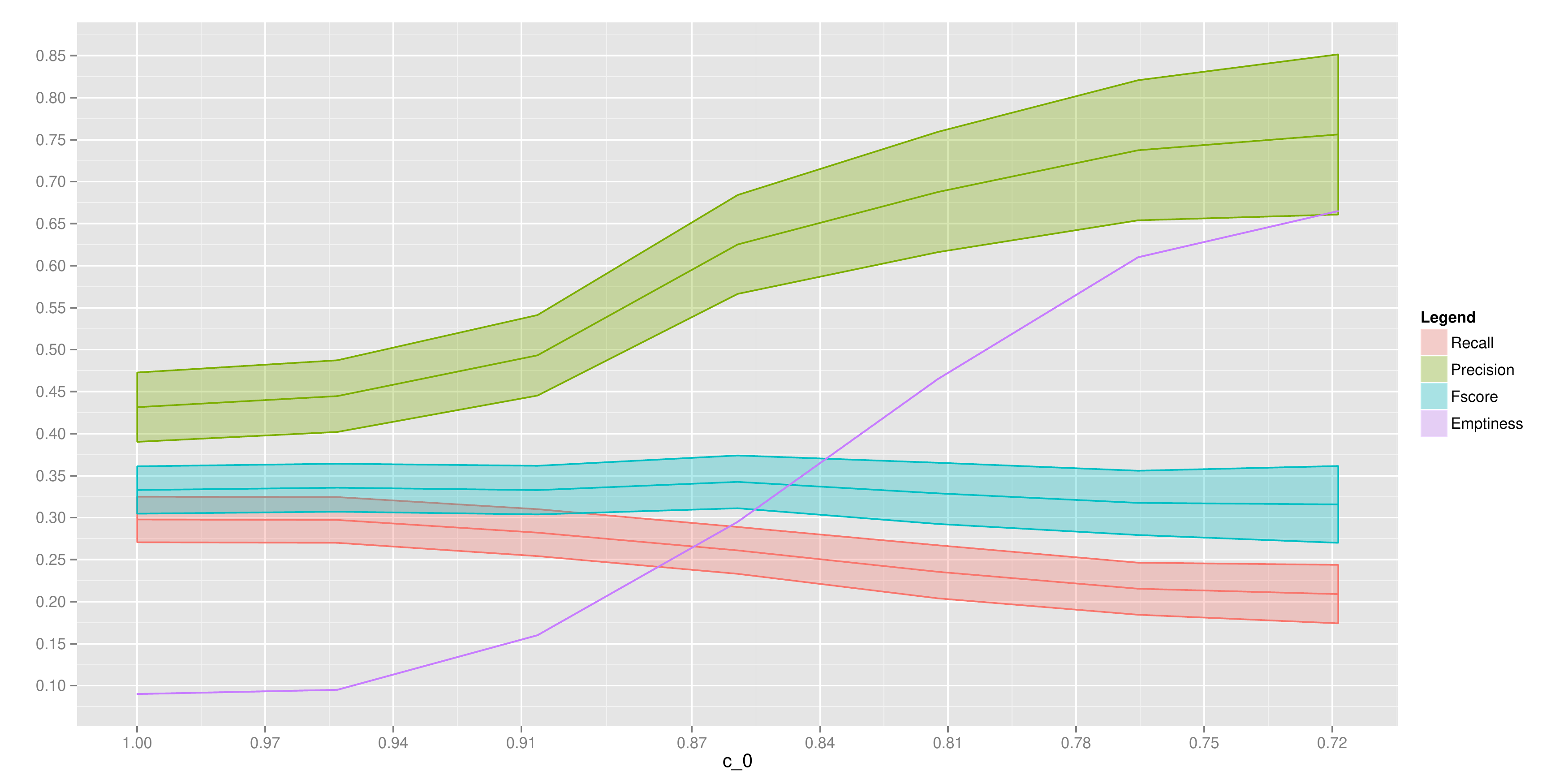}\\
\includegraphics[width=\textwidth]{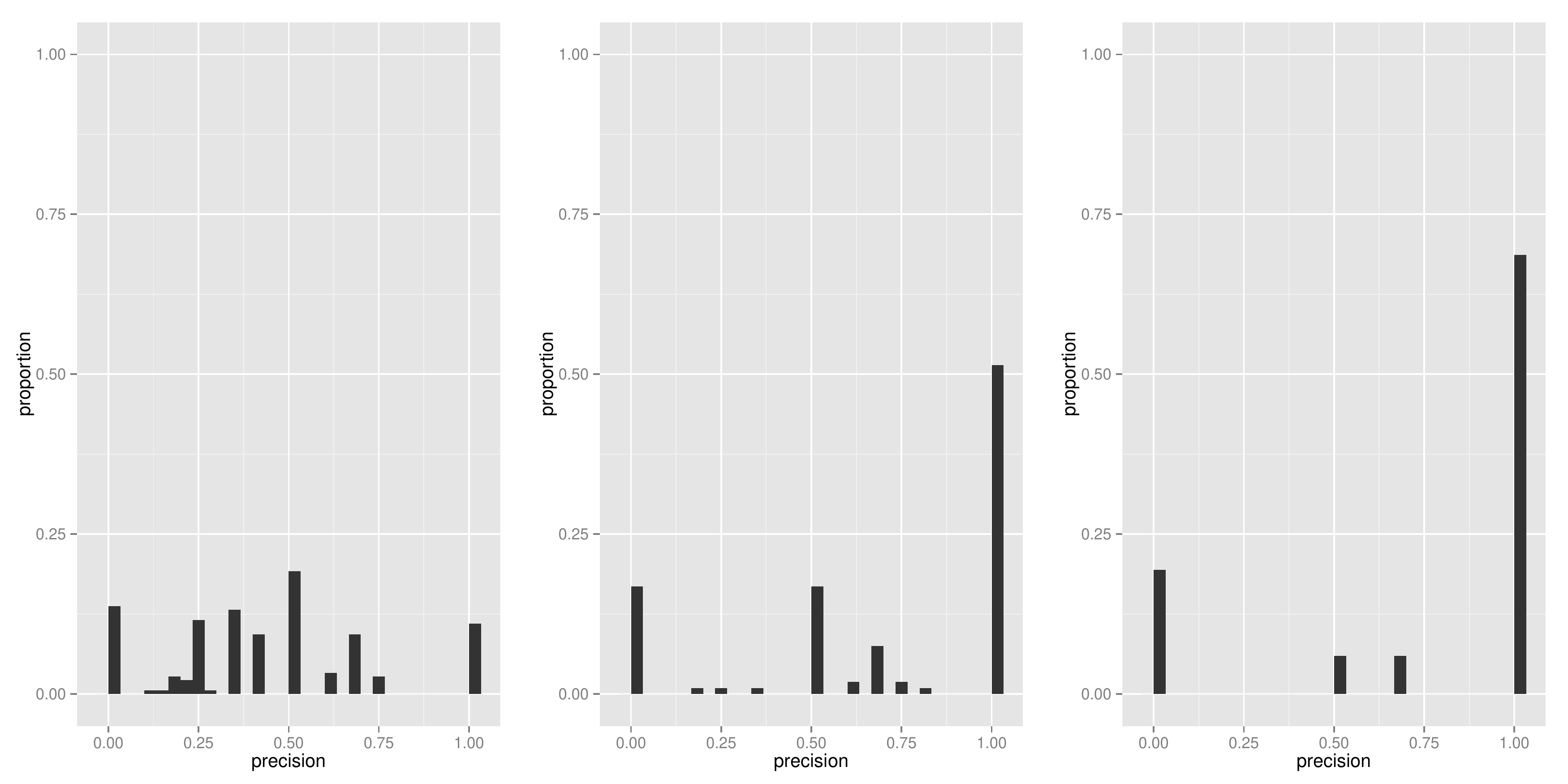}
\end{tabular}
\caption{Situation 4. The histograms show the evolution of the distribution of the precision. From left to right: $c_0=1,0.87,0.72$}
\label{st4}
\end{figure}
\end{document}